%% file: aps.tex
\pdfoutput=1

\documentclass[aps, pre, print, superscriptaddress, twocolumn]{revtex4-1}

\usepackage{amssymb, amsmath}
\usepackage{bm}
\usepackage[bookmarks=false]{hyperref}
\usepackage{graphicx}

\input{00_command.tex}

\begin{document}

\preprint{Ver. 5.0}

\title{Balancing specificity, sensitivity and speed of ligand discrimination by zero-order ultraspecificity}


\author{Masashi K. Kajita}
\email[]{mkajita@sat.t.u-tokyo.ac.jp}
\affiliation{Department of Mathematical Informatics, Graduate School of Information Science and Technology, The University of Tokyo, Japan}

\author{Kazuyuki Aihara}
\affiliation{Department of Mathematical Informatics, Graduate School of Information Science and Technology, The University of Tokyo, Japan}
\affiliation{Institute of Industrial Science, The University of Tokyo, Japan}

\author{Tetsuya J. Kobayashi}
\affiliation{Department of Mathematical Informatics, Graduate School of Information Science and Technology, The University of Tokyo, Japan}
\affiliation{Institute of Industrial Science, The University of Tokyo, Japan}


\date{\today}
\begin{abstract}
\input{abstract.tex}
\end{abstract}

\maketitle


\input{introduction.tex}
\input{results.tex}

\input{discussion.tex}

\begin{acknowledgments}
We thank Omer Dushek for helpful discussions.
This work was partially supported by a Grant-in-Aid for JSPS Fellows (JSPS KAKENHI Grant Number 269545), JSPS KAKENHI Grant Numbers JP15K14433 and 15H05707, and the Platform Project for Supporting in Drug Discovery and Life Science Research (Platform for Dynamic Approaches to Living System) from the Ministry of Education, Culture, Sports, Science  and Technology (MEXT) and Japan Agency for Medical Research and Development (AMED).
\end{acknowledgments}


\bibliographystyle{apsrev4-1}
\bibliography{reference2}

\end{document}

%% file: 00_command.tex
\newcommand{\etal}{\textit{et al.\,}}
\newcommand{\fig}{Fig.\,}
\newcommand{\figs}{Figs.\,}
\newcommand{\Fig}{Figure\,}

\newcommand{\Eq}{Equation\,}
\newcommand{\Eqs}{Equations\,}
\newcommand{\Ref}{Ref.\,}
\newcommand{\tab}{Table\,}
\newcommand{\eq}{Eq.\,}
\newcommand{\eqs}{Eqs.\,}
\newcommand{\diff}{{\rm d}}
\newcommand{\ddf }[1]{\frac{\diff #1}{\diff t}}

\usepackage{ulem}

\newcommand{\delb}[1]{}
\newcommand{\delkj}[1]{}
\newcommand{\addkj}[1]{#1}
\newcommand{\delkjb}[1]{}

\newcommand{\mkp}{\text{MP}}
\newcommand{\gp}{\text{GP}}
\newcommand{\zp}{\text{ZP}}
\newcommand{\cp}{\text{CP}}
\newcommand{\pd}{\text{PdPc}}
\newcommand{\pdpc}{\text{PdPc}}

\newcommand{\X}{X}
\newcommand{\Xtot}{X_{\rm T}}
\newcommand{\R}{R}
\newcommand{\Rp}{R_{\rm p}}
\newcommand{\Rtot}{R_{\rm T}}
\newcommand{\Li}{L}
\newcommand{\Ltot}{L_{\rm T}}
\newcommand{\Ph}{P}
\newcommand{\Ptot}{P_{\rm T}}

\newcommand{\Co}[1]{C_{#1}}
\newcommand{\D}{D}

\newcommand{\Rpstr}{{\overline R_{\rm p}^{*}}}
\newcommand{\Rpr}{{R}_{\rm p}^{*}}

\newcommand{\tCo}[1]{{\rm C}_{#1}}

\newcommand{\F}[1]{\mathcal{F}_{\rm #1}(\R, \Rp)}
\newcommand{\G}[1]{\mathcal{G}_{\rm #1}(\R, \Rp)}

\newcommand{\as}[1]{\mathcal{A}_{#1}}
\newcommand{\bs}[1]{\mathcal{B}_{#1}}
\newcommand{\cs}[1]{\mathcal{C}_{#1}}

\newcommand{\ti}[1]{\tilde{#1}}

\newcommand{\koff}{k_{-1}}
\newcommand{\kv}{w}


%% file: abstract.tex
Specific interactions between receptors and their target ligands in the presence of non-target ligands are crucial for biological processes such as T cell ligand discrimination.
To discriminate between the target and non-target ligands, cells have to increase specificity by amplifying the small differences in affinity among ligands. 
In addition, sensitivity to ligand concentration and quick discrimination are also important to detect low amounts of target ligands and facilitate fast cellular decision making after ligand recognition. 
In this work, we find that ultraspecificity is naturally derived from a well-known mechanism for zero-order ultrasensitivity to concentration. 
We also show that this mechanism can produce an optimal balance of specificity, sensitivity, and quick discrimination.
Furthermore, we show that a model for insensitivity to large number of non-terget ligands can be naturally derived from the ultraspecificity model.
Zero-order ultraspecificity may provide alternative way to understand ligand discrimination from the viewpoint of the nonlinear properties of biochemical reactions.

%% file: introduction.tex
\section{Introduction}
Cellular systems rely on various biochemical reactions for signal transduction, genetic information processing, or metabolism.
Signal transduction systems, for example, have receptors to detect their specific (target) ligands.
After binding of the ligands to the receptors, the receptors transmit information about the existence and the concentration of the ligand to the inside of the cells.
Based on that information, the cells make appropriate decisions such as gene expression, cell division, chemotaxis, and the immune response.
However, intracellular and extracellular systems consist of various molecules, most of which are non-specific (non-target) molecules of the receptors.
Therefore, the non-target molecules also have chances to bind with the receptors.
If the receptors were to bind with the non-target ligands and send incorrect information about the detections of the target ligands, it might cause serious problems for the cell.
To suppress such incorrect reactions, biochemical reactions need to discriminate between their target and non-target molecules.

Among various kinds of cells, T cell, a type of immune cells, is known to have an efficient discriminative ability.
T cells have T cell receptor (TCR) on their surface to recognize ligand molecules presented by antigen-presenting cells.
If T cells respond to non-target ligands, it may cause an inappropriate immune response such as inflammation.
Therefore, it is natural for T cells to attain the ability to respond only to their target  ligands.
Such  selective response to target ligands is known as ``specificity''.
It has been demonstrated experimentally that T cells can discriminate their target from non-target ligands, even when a non-target ligand has only a single amino acid substitution of a target ligand \cite{1998Natur.395...82M, JanewayDJnhpEb, 1991Sci...252.1308E}.
In addition to specificity, recent quantitative experiments have also elucidated that T cells possess three additional properties relevant for ligand discrimination: sensitivity to a low amount of the target ligands, quick discrimination, and insensitivity to a large amount of non-target ligands \cite{2016JSP...162.1130F, Kajita:2015eh}.

Sensitivity is an ability to generate an all-or-none response to a small change around threshold in the target ligand concentration, and it has been confirmed experimentally that T cells can respond to fewer than ten target ligand molecules \cite{Huang:2013jn, 2002Natur.419..845I}.
Quick discrimination literally means that a cell can respond quickly to target ligands. 
Actually, T cell activation in response to target ligands starts within 1 -- 5 minutes after the presentation of the target ligands \cite{2002Natur.419..845I, Yokosuka2005}.
Sensitive and a prompt response are necessary properties for an appropriate immune response at the early stage of infection when there exists only a small amount of target non-self ligands.
A third required property is concentration compensation to generate insensitivity to the non-target ligands at their physiological concentrations.
Even for a receptor that has a high specificity to the target ligands, a high concentration of non-target ligands may induce the activation of the receptor, especially when the receptor is highly sensitive to a small amount of the target ligands.
Therefore, T cells must realize insensitivity to non-target ligands by compensating for a high concentration of the non-target ligands without losing the sensitivity to their target ligands.
Evavold \etal reported that a high concentration of an altered peptide ligand with a single amino acid substitution did not induce T cell activation, namely T cell proliferation and interleukin-4 production, whereas a low concentration of the original peptide induced T cell activation \cite{1991Sci...252.1308E}.
Having sensitivity to target ligands at a low concentration and insensitivity to non-target ligands at a high concentration seems contradictory.
However, quantitative experiments on the four properties of cell surface ligand-binding receptors (specificity \cite{1998Natur.395...82M, JanewayDJnhpEb,1991Sci...252.1308E}, sensitivity to target \cite{Huang:2013jn, 2002Natur.419..845I}, quick discrimination \cite{2002Natur.419..845I, Yokosuka2005}, and insensitivity to non-target \cite{1991Sci...252.1308E}) suggest that T cells are equipped with mechanisms to balance all of these four properties for accurate ligand discrimination in various situations.

To understand the underlying mechanisms of ligand discrimination, the relationship between 
nonlinear responses to ligands and  reaction network structures has been investigated.
Hopfield and Ninio independently proposed the existence of a ``kinetic proofreading'' mechanism as an explanation for the high specificity \cite{1974PNAS...71.4135H, Ninio1975}.
This mechanism was originally proposed to explain the remarkable fidelity of protein synthesis and DNA replication. 
McKeithan applied the idea for T cell ligand discrimination \cite{1995PNAS...92.5042M}.
In the kinetic proofreading, a non-equilibrium irreversible reaction is crucial to enhance specificity beyond the simple difference in the affinity of receptors to the target and non-target ligands.
Because the single discrimination reaction was not enough to explain the high specificity observed in biological processes, they extended the single proofreading step model to a sequentially connected multistep proofreading model to exponentially amplify the small differences in affinity.
While the multistep kinetic proofreading  amplifies specificity,  the multiple reactions are accompanied by a loss of sensitivity \cite{George2005} as well as a time delay.
To balance specificity and sensitivity, several modifications have been proposed, e.g., by extending the multistep scheme \cite{2012PNAS..10912034M} or by incorporating feedback loops \cite{Lever2014, Altanbonnet:2005hoa, 2008Sci...321.1081F}.
In addition, Lalanne \etal applied \textit{in silico} evolution to obtain an ``adaptive sorting'' model \cite{2016JSP...162.1130F, 2013PhRvL.110u8102L}
that balances the specificity, sensitivity to target, and insensitivity to non-target.
The adaptive sorting model is a modified version of a multistep kinetic proofreading model with a feedback regulation motif, which is similar to the bacterial two-component system for initiating an adaptive response to an external signal \cite{1999Natur.397..168A}.
All of the theoretical research to date has suggested that nonlinear regulation can resolve the trade-off between specificity and sensitivity. 
However, this trade-off may be resolved by considering different kinetics than those of multistep kinetic proofreading for specificity because the trade-off originates from this multistep process. 
Compared with modifications of multistep proofreading, this possibility has not yet been fully investigated.

In this work, we investigate zero-order ultrasensitivity as an alternative mechanism for enhancing specificity.
Zero-order ultrasensitivity was originally proposed as a mechanism for ultrasensitivity to a change in concentration of the ligand \cite{1981PNAS...78.6840G}, in which ultrasensitivity  is realized by zero-order saturated reactions rather than nonlinear cascading reactions.
The saturation condition works as a zero-order reaction.
In \Ref\cite{2003BpJ....84.1410Q}, Qian analyzed the discrimination efficiencies of the original model of zero-order ultrasensitivity  proposed in \Ref\cite{1981PNAS...78.6840G}.
However, in this case, specificity is just a byproduct of ultrasensitivity. 
By adopting a zero-order property directly to enhance specificity, high specificity by a zero-order reaction mechanism, denoted in this work as "zero-order specificity", may be obtained.
However, the discrimination efficiencies of the combination of the kinetic proofreading and zero-order ultrasensitivity mechanisms have not yet been investigated \cite{2003BpJ....84.1410Q}.
In this paper, we combine the zero-order ultrasensitivity mechanism with single-step kinetic proofreading to show how the mechanism can simultaneously accommodate specificity, sensitivity, and speed.
We also demonstrate that a model with insensitivity to non-target ligands can be naturally derived from our zero-order ultraspecificity model.

The rest of the paper is organized as follows: we first introduce a multiple kinetic proofreading model with a phosphorylation and dephosphorylation cycle. 
Then we characterize the properties of the model such as specificity, sensitivity and speed.
Next, we introduce our model of zero-order ultraspecificity and characterize its properties.
To find an optimal combination of multiple proofreading steps and the zero-order ultraspecificity for ligand discrimination,
we also introduce a generalized kinetic proofreading model that has multiple kinetic proofreading reactions with zero-order reactions, and analyze the effects of the multiple reactions and zero-order reactions on the discrimination properties.
Furthermore, we derive a ligand concentration compensation model from the zero-order ultraspecificity model and analyze its discrimination efficiencies.
Finally, we summarize the relationship among several kinetics models and their properties for ligand discrimination. 
We also discuss the biological relevance of the mechanisms and future directions by focusing on T cell ligand discrimination.

%% file: results.tex
\section{Zero-order ultrasensitivity for ligand discrimination}

\subsection{Modeling ligand discrimination}

\begin{figure}[htbp]
  \begin{center}
   \includegraphics[width=0.45\textwidth]{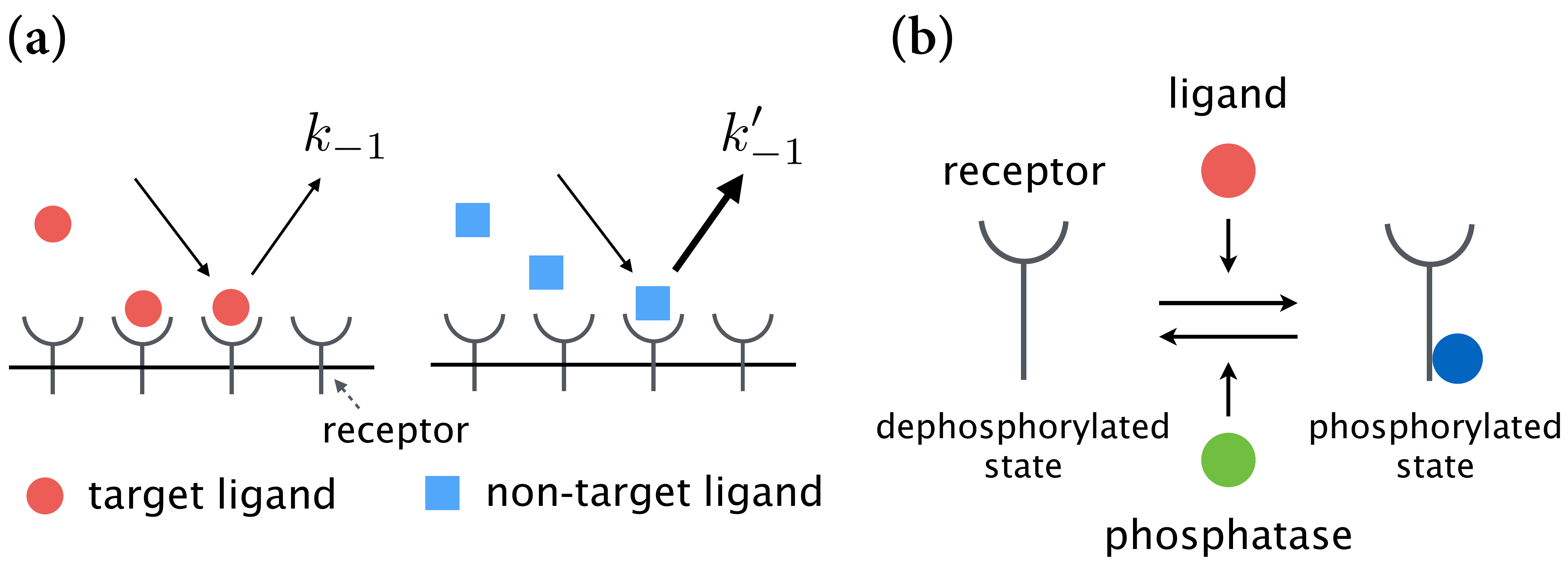}
   \hspace{1.6cm} 
    \caption{
(a) A schematic diagram of ligand binding to receptors. 
Target and non-target ligands have different unbinding rates $k_{-1}$ and $k_{-1}^{\prime}$. 
The unbinding rate of the target ligand is smaller than that of the non-target ligand as $k_{-1} < k_{-1}^{\prime}$. 
(b) Receptor phosphorylation and dephosphorylation reactions in a ligand discrimination system. 
Receptor phosphorylation is induced by ligand whereas receptor dephosphorylation is catalyzed by phosphatase enzymes.
}
  \label{fig:1}
  \end{center}
\end{figure}

The simplest ligand discrimination process by a cell is a binary discrimination in which the cell determines whether the ligand bound to the receptors is their target or not. 
We assume that cells discriminate target from non-target ligands based on their affinity to receptors, more precisely, their unbinding rates $k_{-1}$ (\fig \ref{fig:1}(a)) \cite{1974PNAS...71.4135H, Ninio1975, 1995PNAS...92.5042M}.

In this study, we model ligand discrimination using a receptor phosphorylation-dephosphorylation cycle.
Some of the previous works have not explicitly modeled the receptor dephosphorylation reaction \cite{1974PNAS...71.4135H, Ninio1975, 1995PNAS...92.5042M, 2016JSP...162.1130F}.
However, in general, phosphorylated molecules are dephosphorylated by phosphatase enzymes. 
Therefore, we employ a phosphorylation-dephosphorylation cycle.
The phosphorylation reaction is induced by a ligand \rm{\Li} bound to a receptor $\rm{\R}$.
The phosphorylated receptor $\rm{\Rp}$ is dephosphorylated by a phosphatase $\rm \Ph$ (\fig \ref{fig:1}(b)). 
We also assume that the discrimination decision is made based on the ratio of the phosphorylated receptors to the total receptors $\Rtot$ as $\Rpr:=\Rp/\Rtot$. 
Because a ligand discrimination system should be activated by its target ligands and should not be activated by non-target ligands, the phosphorylation ratio $\Rpr$ should be high when the system has the target ligands and $\Rpr$ should be low when the system has non-target ligands.

\subsection{Kinetic proofreading with multiple proofreading reactions}

Next, we introduce a multiple kinetic proofreading (\mkp) model to evaluate the effects of a multistep reaction on the discrimination properties. 
$\mkp$ kinetics is basically the same as the kinetics described in \Ref\cite{1974PNAS...71.4135H, 1995PNAS...92.5042M} but extended to have a spontaneous dephosphorylation reaction (\eq \ref{model:MP}).
\begin{equation}\label{model:MP}
\includegraphics[width=0.42\textwidth]{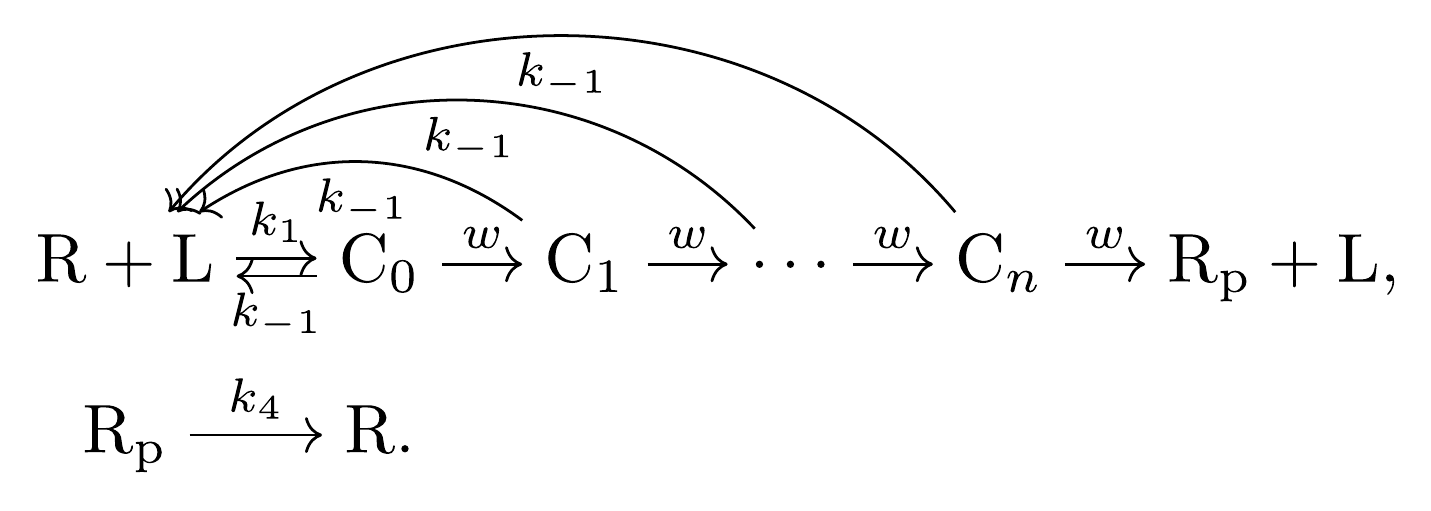}
\end{equation}
Ligand $\rm{\Li}$ binds to $\rm \R$ with rate $k_{1}$ to form intermediate complexes, $\tCo{i}$ (for $i=0, 1, ..., n$), before reaching the phosphorylated receptor $\rm \Rp$. 
The reactions from $\tCo{i-1}$ to $\tCo{i}$ (for $i=1, ..., n$) are irreversible and proceed at rate $\kv$.  
Ligand $\rm{\Li}$ can dissociate from the receptor at any intermediate complex state with unbinding rate $k_{-1}$, returing the receptor to the dephosphorylated state $\rm{\R}$. 
The phosphorylated receptor $\rm \Rp$ is dephosphorylated according to first-order reaction kinetics with rate $k_{4}$. 
We assume that the kinase for the phosphorylation reaction and the phosphatase for the dephosphorylation reaction are present at constant levels so that the concentration of the enzymes can be included in the reaction constants.

By assuming that each reaction follows the law of mass action, the dynamics of $\mkp$ kinetics defined by \eq \ref{model:MP} can be described by the following ordinary differential equations (ODEs):
\begin{eqnarray}
&& \ddf \R = - k_{1} \R \Li + k_{4} \Rp + k_{-1} \sum_{i=0}^{n}{\Co{i}},\label{eq:MP-1}\\
&& \ddf \Li = - k_{1} \R \Li + \kv {\Co{n}} + k_{-1} \sum_{i=0}^{n}{\Co{i}},\label{eq:MP-2} \\
&& \ddf \Rp = \kv {\Co{n}} - k_{4} \Rp,\label{eq:MP-3}\\
&& \ddf{\Co{0}} = - k_{-1} {\Co{0}} - \kv {\Co{0}} + k_{1} \R \Li,\label{eq:MP-4}\\
&& \ddf{\Co{i}} = - k_{-1} {\Co{i}} - \kv {\Co{i}} + \kv {\Co{i-1}}, \, (i=1, ..., n), \label{eq:MP-5}
\end{eqnarray}
where $\R$, $\Li$, $\Rp$, and $\Co{i}$ (for $i=0,1,\dots, n$) are the concentrations of $\rm \R$, $\rm \Li$, $\rm \Rp$, and $\tCo{i}$ (for $i=0,1,\dots, n$), respectively.

For mathematical simplicity, we approximate \eqs \eqref{eq:MP-1} -- \eqref{eq:MP-5} with a quasi-steady-state assumption such that $\diff {\Co{i}}/ \diff t = 0$ for $i=0,1, \dots, n$.
This approximation means that $\R$ and $\Rp$ change slowly compared with $\Li$ and $\Co{i}$ (for $i=0,1, \dots, n$). 
Then we obtain
\begin{eqnarray}
&&\frac{\diff \R}{\diff t}
 =  \G{\mkp} - \F{\mkp} \label{eq:MP-qss-R},\\
&&\frac{\diff \Rp}{\diff t}
 =  \F{\mkp} - \G{\mkp}, \label{eq:MP-qss}
\end{eqnarray}
where $\F{\mkp}:=V_{1} {\R}/{(K_{m1}+\R)}$ and $\G{\mkp}:=k_{4} \Rp$, $V_{1}:=\kv \Ltot {\alpha^{n}}/{\sum_{i=0}^{n}\alpha^{i}}$, $K_{m1}:=K_{1}/\sum_{i=0}^{n}\alpha^{i}$, and $K_{1}:=(k_{-1}+\kv)/k_{1}$ (see also \tab \ref{tab:table1}).
Here, $K_{m1}$ is the Michaelis--Menten constant, 
$V_{1}$ is the maximum velocity of the Michaelis--Menten reaction, 
$\alpha:=\kv/(k_{-1}+\kv)$ is the probability that a ligand-receptor complex will be modified before the complex dissociates,
and $\Ltot:=\Li+\sum_{i=0}^{n}\Co{i}$ is the total concentration of ligand.

Because $\R + \Rp$ is constant over time as $\diff \R/\diff t + \diff \Rp/\diff t =0$ from \eqs \eqref{eq:MP-qss-R} -- \eqref{eq:MP-qss}, we focus only on \Eq \eqref{eq:MP-qss}. 
At the steady-state, $\diff \Rp/\diff t=0$, the ratio of phosphorylated receptor according to $\mkp$ kinetics can be derived as
\begin{eqnarray} \label{eq:MP-Rpstr}
\Rpstr= 
\frac{\bs{\mkp} - \sqrt{ \bs{\mkp}^{2} - 4 \as{\mkp} \cs{\mkp} }}{2 \as{\mkp}},
\end{eqnarray}
for $0 \leq \Rpr \leq 1$, and $\Rpstr$ is the steady-state value of $\Rpr$.
Here $\as{\mkp}:=k_{4}$, $\bs{\mkp}:=V_{1}/\Rtot + k_{4}( 1 + K_{m1} / \Rtot)$, and $\cs{\mkp}:=V_{1}/\Rtot$, where $\Rtot:=\R+\Rp+\sum_{i=0}^{n}\Co{i}$.
For this derivation, we introduce the assumption that the total receptor concentration $\Rtot$ is much larger than the total ligand concentration $\Ltot$, that is, $\Rtot \gg \Ltot$.

\begin{figure*}[htbp]
  \begin{center}
    \includegraphics[width=\textwidth]{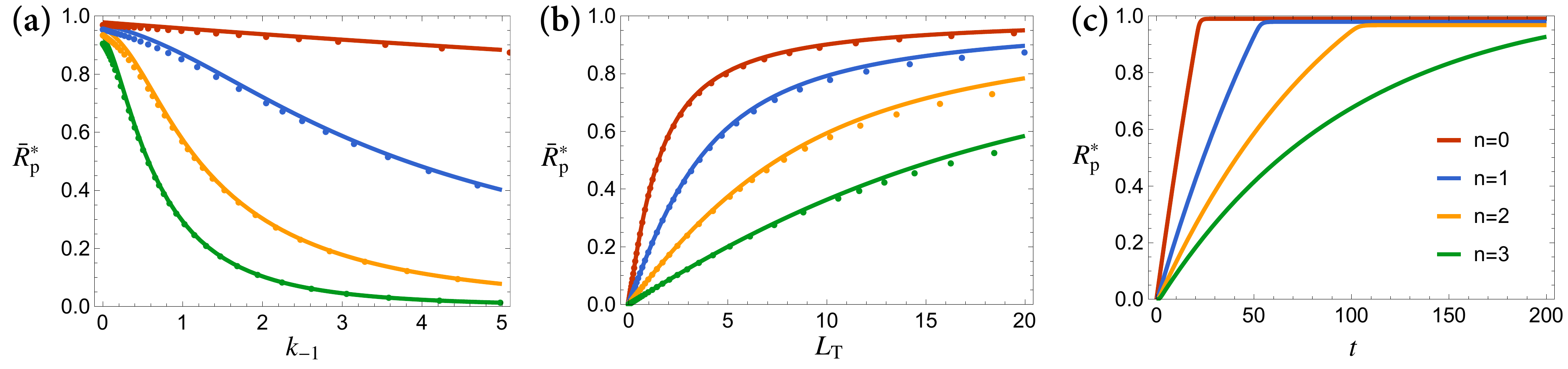}
    \hspace{1.6cm} 
    \caption{
The steady-state response and the time series of the $\mkp$ kinetics for various numbers of proofreading steps $n$: $n=0$ (red), $n=1$ (blue), $n=2$ (yellow), and $n=3$ (green).
(a, b)
The stationary fraction of the phosphorylated receptor $\Rpstr$ as a function of the unbinding rate $\koff$ with $\Ltot=5$ and $k_{1}=0.11$ (a), and $\Rpstr$ as a function of the total ligand concentration $\Ltot$ with $\koff=1$ and $k_{1}=0.02$ (b). 
The solid curves are analytically obtained from \eq \eqref{eq:MP-Rpstr} and the dotted curves are obtained from the numerical simulation. 
(c) 
The time series of the fraction of phosphorylated receptor $\Rpr(t)$ with $\Ltot=5$, $\koff=0.1$, and $k_{1}=10$. 
In (a) - (c), the values of the other parameters are $\Rtot=100$, $\kv=1$, and $k_{4}=0.01$.
The results of the numerical simulations are obtained from \eqs \eqref{eq:MP-1} -- \eqref{eq:MP-5}. 
We use the same initial values, $\R(t=0)=\Rtot$, $\Rp(t=0)=0$, $\Co{i}(t=0)=0$ (for $i=0, 1, \dots, n$), and $\Li(t=0)=\Ltot$.
}
    \label{fig:MP2_Rpr()_n}
  \end{center}
\end{figure*}

To understand the responses of the $\mkp$ kinetics to the unbinding rate $\koff$ and to the total ligand concentration $\Ltot$, we calculate the steady state of the $\mkp$ kinetics as a function of $\koff$ and $\Ltot$.
\Fig \ref{fig:MP2_Rpr()_n}(a) is the steady state response of the phosphorylated receptor fraction $\Rpstr(k_{-1})$ as a function of the unbinding rate $k_{-1}$. 
As $n$ increases, the nonlinearity of $\Rpstr(k_{-1})$ increases.
This nonlinearity enables the $\mkp$ kinetics with a large $n$ to discriminate target ligands from non-target ligands.
For example, when $\koff$ is close to $0.5$, $\Rpstr(k_{-1})$ with a large $n$ is greatly changed by small changes in $\koff$. 
Thus, it can sharply discriminate target from non-target ligands even if they have similar $\koff$ values.
However, $\Rpstr(k_{-1})$ with a small $n$ is changed only slightly by small changes in $\koff$; therefore, it cannot discriminate the target from non-target ligands.
We also plot the fraction of phosphorylated receptor $\Rpstr$ as a function of the total ligand concentration $\Ltot$ as in \fig \ref{fig:MP2_Rpr()_n}(b).
As $n$ increases, the nonlinearity of $\Rpstr(\Ltot)$ decreases.
The weak nonlinear response means that the $\mkp$ kinetics with large $n$ is not sensitive to ligand concentration.
For example, when $\Ltot$ is close to $0$, $\Rpstr(\Ltot)$ with a small $n$ is greatly changed by small changes in $\Ltot$, and then it can detect and respond to small amounts of the ligand.
However, $\Rpstr(k_{-1})$ with a large $n$ is slightly changed by small changes in $\Ltot$; therefore, it cannot detect small amounts of the ligand.
Overall, in $\mkp$ kinetics, the number of proofreading steps $n$ increases responsiveness to $\koff$ (specificity), whereas it reduces responsiveness to $\Ltot$ (sensitivity).
This result indicates that there is a trade-off between specificity and sensitivity in $\mkp$ kinetics if we change $n$.
Finally, we evaluate speed of response as in \fig \ref{fig:MP2_Rpr()_n}(c) by plotting $\Rpr(t)$ for $n=0, 1, 2$, and $3$. 
The increase in $n$ elongates the time to reach the stationary state. 

These results demonstrate that, in the $\mkp$ kinetics, having multiple proofreading steps increases the nonlinearity of the stationary response to the unbinding rate $\koff$, and therefore the multiple steps amplify specificity (\fig \ref{fig:MP2_Rpr()_n} (a)). 
However, the multiple steps amplify specificity at the cost of a loss of sensitivity, a quick response to changes in ligand concentration, and speed (\figs \ref{fig:MP2_Rpr()_n}(b, c)) (see also \tab \ref{tab:table1}). 

\subsection{Kinetic proofreading with zero-order ultrasensitivity}

Next, we consider ligand discrimination by the zero-order ultrasensitivity mechanism. 
The Goldbeter--Koshland $\pd$ kinetics \cite{1981PNAS...78.6840G} shown in \eq \ref{model:PdPc} has ultrasensitivity to concentration in saturated conditions.
In the $\pd$ kinetics, both the phosphorylation and the dephosphorylation reactions are Michaelis--Menten reactions whose substrates, $\rm \R$ or $\rm \Rp$, can be catalyzed by $\rm \Li$ or $\rm \Ph$ after the formation of the complex $\tCo{0}$ or $\rm \D$ as
\begin{equation}\label{model:PdPc}
\includegraphics[width=0.25\textwidth]{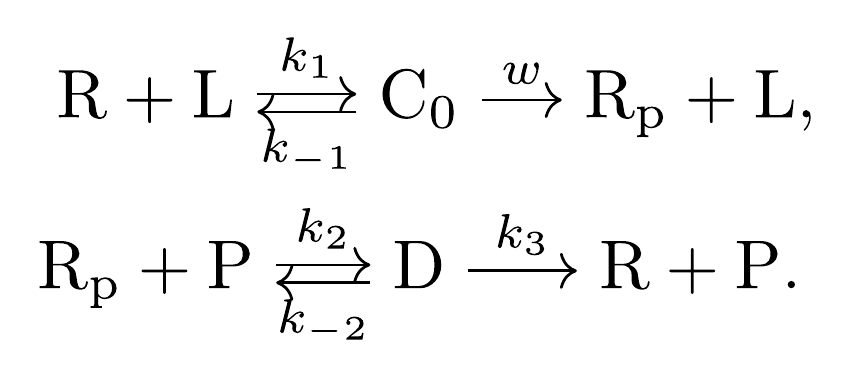}
\end{equation}
Here, $k_{2}$, $k_{-2}$, and $k_{3}$ are the reaction constants.

To derive a model including the zero-order specificity mechanism, we combine the kinetic proofreading mechanism with the zero-order ultrasensitivity mechanism by modifying the $\pd$ kinetics (\eq \ref{model:PdPc}) to have a single proofreading reaction in the phosphorylation reaction as shown in \eq \ref{model:ZP}. 
We denote the kinetics shown in \eq \ref{model:ZP} as zero-order proofreading (\zp) kinetics:
\begin{equation}\label{model:ZP}
\includegraphics[width=0.35\textwidth]{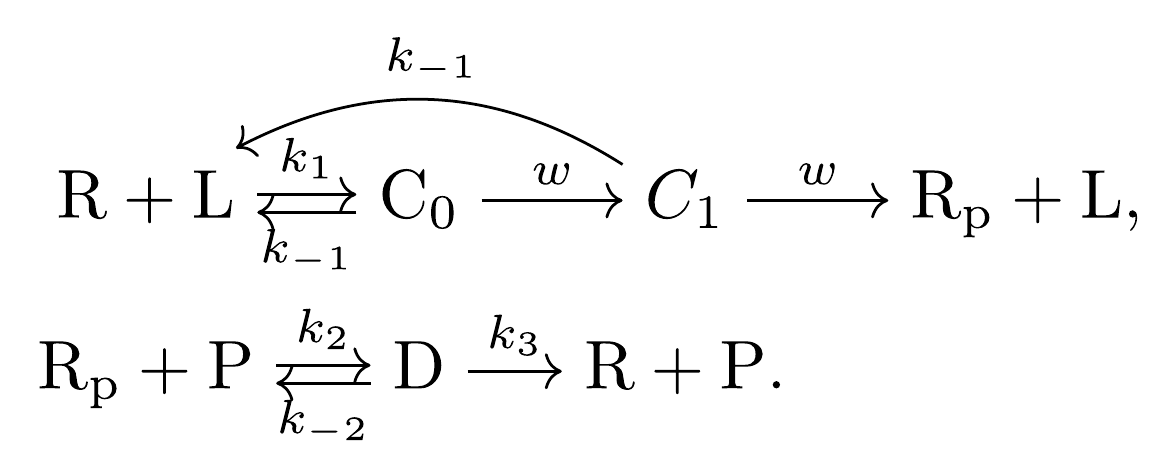}
\end{equation}
We can extend $\zp$ kinetics to have multiple proofreading steps as shown in \eq \ref{model:GP}. 
We denote it as a generalized kinetic proofreading (\gp) model: 
\begin{equation}\label{model:GP}
\includegraphics[width=0.46\textwidth]{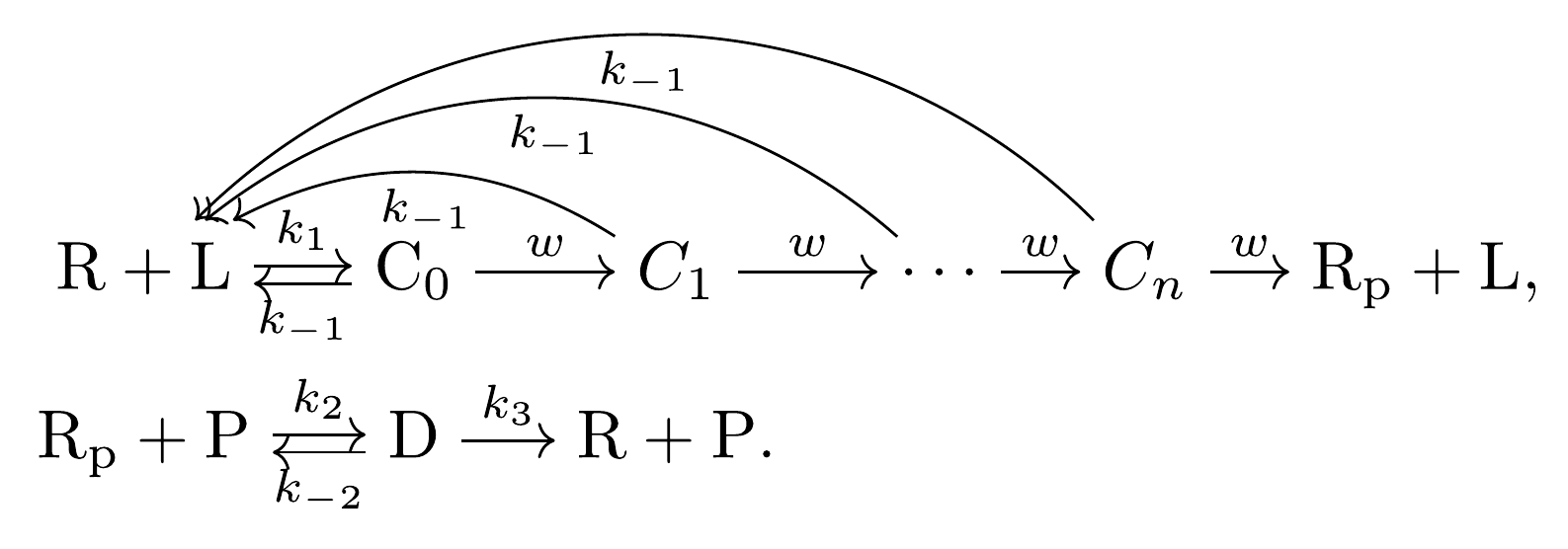}
\end{equation}

The dynamics of the $\gp$ model, which includes both the original $\pd$ kinetics ($n=0$) and $\zp$ kinetics ($n=1$), can be represented by the following ODEs:
\begin{eqnarray}\label{eq:ZP}
&& \ddf \R = - k_{1} \R \Li + k_{3} \D + k_{-1} \sum_{i=0}^{n}{\Co{i}},\label{eq:ZP-1}\\
&&\ddf \Li = - k_{1} \R \Li + \kv \Co{n} + k_{-1} \sum_{i=0}^{n}{\Co{i}},\label{eq:ZP-2}\\
&&\ddf \Rp =\kv {\Co{n}} - k_{2} \Rp \Ph + k_{-2} \D,\label{eq:ZP-3}\\
&&\ddf \Ph = -k_{2} \Rp \Ph + k_{-2} \D + k_{3} \D,\label{eq:ZP-4}\\
&&\ddf \D = k_{2} \Rp \Ph -k_{-2} \D - k_{3} \D,\label{eq:ZP-5}\\
&&\ddf {\Co{0}} = -k_{-1} \Co{0} - \kv \Co{0} + k_{1} \R \Li,\label{eq:ZP-6}\\
&&\ddf {\Co{i}} = -k_{-1} \Co{i} - \kv \Co{i} + \kv \Co{i-1}, (i=1, 2, \dots, n), \label{eq:ZP-7}
\end{eqnarray}
where $\Ph$ and $\D$ are the amounts of $\rm{\Ph}$ and $\rm{\D}$, respectively. 
When $n=0$, \eqs \eqref{eq:ZP} -- \eqref{eq:ZP-7} can be reduced to those of $\pd$ kinetics (\eq \ref{model:PdPc}). 

As for $\mkp$ kinetics, we approximate the $\gp$ model (\eqs \eqref{eq:ZP-1} -- \eqref{eq:ZP-7}) by assuming the quasi-steady-state condition such that $\diff {\Co{i}}/ \diff t = 0$ (for $i=0,1, \dots, n$) and $\diff \D/\diff t=0$. 
Then we obtain
\begin{eqnarray}\label{eq:gp-qss}
&&\frac{\diff \R}{\diff t}
 =  \G{\gp} - \F{\gp}, \label{eq:zp-qss-1}\\
 &&\frac{\diff \Rp}{\diff t}
 =  \F{\gp} - \G{\gp}, \label{eq:zp-qss-2}
\end{eqnarray}
where $\F{\gp}:=V_{1} {\R}/{(K_{m1}+\R)}$ and $\G{\gp}:=V_{2}{\Rp}/{(K_{2}+\Rp)}$ (see also \tab \ref{tab:table1}). 
Here, 
$V_{1}=\kv \Ltot {\alpha^{n}}/{\sum_{i=0}^{n}\alpha^{i}}$ and $V_{2}=k_{3} \Ptot$ are the maximum velocities of the Michaelis--Menten reactions, respectively.
In addition, $K_{m1}=K_{1}/\sum_{i=0}^{n}\alpha^{i}$ and $K_{2}:=(k_{-2}+k_{3})/k_{2}$ are the Michaelis--Menten constants, $\Ltot=\Li+\sum_{i=0}^{n}\Co{i}$ is the total ligand concentration, and $\Ptot:=\Ph+\D$ is the total phosphatase concentration.
Because $\diff R/\diff t + \diff \Rp/\diff t =0$, we focus only on \eq \eqref{eq:zp-qss-2}. 

Note that the $\gp$ model also includes $\mkp$ kinetics as a special case.
When we compare \eqs \eqref{eq:MP-qss} and \eqref{eq:zp-qss-2}, $\F{\mkp}=\F{\gp}$ but $\G{\mkp} \neq \G{\gp}$. 
The difference between $\G{\mkp}$ and $\G{\gp}$ is caused by the difference between the dephosphorylation reactions in $\mkp$ kinetics and $\gp$ kinetics. 
The $\mkp$ kinetics has a first-order dephosphorylation reaction whereas the $\gp$ model has a Michaelis--Menten type reaction. 
The Michaelis--Menten equation is a nonlinear function of $\Rp$. 
However, it can be approximated to be a linear function of $\Rp$ in the following case:
When $K_{2}$ of $ \G{\gp}$ is much larger than $\Rp$, then $k_{3} \Ptot {\Rp}/{(K_{2}+\Rp)} \approx k_{3} \Ptot \Rp / K_{2}$. 
This relationship is always valid if we assume $K_{2} \gg \Rtot \geq \Rp$ where $\Rtot =\R + \Rp + \sum_{i}^{n} \Co{i}$. 
Therefore, when $k_{3} \Ptot / K_{2} = k_{4}$ holds, the $\gp$ model is approximately equivalent to the $\mkp$ kinetics, $\G{\mkp} \approx \G{\gp}$.

\subsubsection{Steady-state response of the ZP kinetics}

The steady-state fraction of the phosphorylated receptor of the $\gp$ model can be derived from \eq \eqref{eq:zp-qss-2} as
\begin{eqnarray} \label{eq:ZPsol1}
\displaystyle 
\Rpstr=\frac{\bs{\gp} - \sqrt{\bs{\gp}^{2} - 4 \as{\gp} \cs{\gp}} }{2 \as{\gp}},
\end{eqnarray}
where $\as{\gp}:=1 - {V_{1}}/{V_{2}}$, $\bs{\gp}:=(1 - {V_{1}}/{V_{2}})+\ti{K_{2}} {V_{1}}/{V_{2}}  + \ti{K_{1}}$, and $\cs{\gp}:=\ti{K_{2}}{V_{1}}/{V_{2}} $.
Here, $\ti{K_{1}}:=K_{m1}/\Rtot$, and $\ti{K_{2}}:=K_{2}/\Rtot$.
For a special case where $V_{1}/V_{2}=1$, we obtain
\begin{eqnarray} \label{eq:ZPsol2}
\Rpstr=\frac{K_{2}}{K_{m1}+K_{2}}.
\end{eqnarray}
For this derivation, we introduce an assumption that the total concentration of receptor $\Rtot$ is much larger than that of ligand $\Ltot$ and that of phosphatase $\Ptot$, that is, $\Rtot \gg \Ltot$ and $\Rtot \gg \Ptot$.

\begin{figure*}[!htbp]
\begin{center}
    \includegraphics[width=0.9\textwidth]{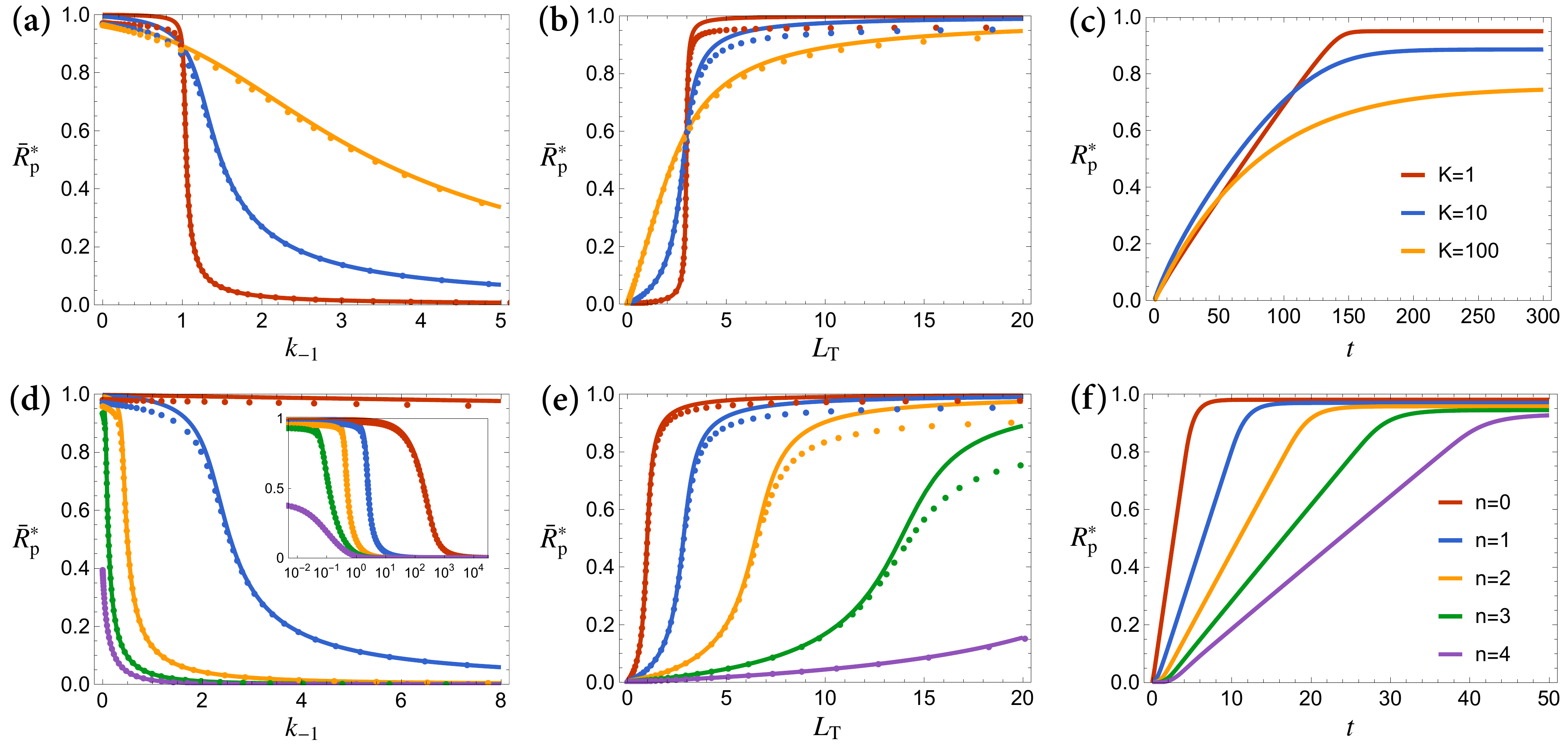}
\caption{
The steady-state response and the time series of the $\gp$ model for different values of the unsaturation level $K$ when the number of the proofreading step is $n=1$ (a -- c), and for different values of the proofreading steps $n$ when $K=10$ (d -- f). 
The $\gp$ model with small $K$ and small $n$ corresponds to the $\zp$ kinetics, and the $\gp$ model with large $K$ and large $n$ corresponds to the $\mkp$ kinetics.
(a, d) The stationary fraction of the phosphorylated receptor $\Rpstr$ as a function of the unbinding rate $\koff$. 
The semilogarithmic plot is also shown in the inset of (d).
(b, e) The stationary fraction of the phosphorylated receptor $\Rpstr$ as a function of the total ligand concentration $\Ltot$.
(c, f) The time series of the fraction of phosphorylated receptor $\Rpr(t)$ obtained from the numerical simulation. 
The curves in (a, b, d, e) are analytically obtained from \eqs \eqref{eq:ZPsol1} and \eqref{eq:ZPsol2}, and the dotted lines are obtained from the numerical simulation.
In (a -- c), the value of the unsaturation level is $K=1$ (red), $K=10$ (blue), and $K=100$ (yellow).
The parameters are $\Ltot=3$ in (a), and $\Ltot=5$ in (c), respectively.
The values of the other parameters are $\Rtot=100$, $\Ptot=1$, $\kv=1$, $k_{-2}=10$, and $k_{3}=1$.
In (d -- f), the number of the proofreading steps is $n=0$ (red), $n=1$ (blue), $n=2$ (yellow), $n=3$ (green), and $n=4$ (purple).
The parameters are $\Ltot=4$ in (d), and $\Ltot=20$ in (f), respectively.
The values of the other parameters are $\Rtot=100$, $\Ptot=1$, $\kv=1$, $k_{-2}=10$, and $k_{3}=1$.
Here $k_{1}$ and $k_{2}$ are obtained as $k_{1}=(\bar{k}_{-1}+\kv)/K_{1}$ and $k_{2}=(k_{-2}+k_{3})/K_{2}$ where $K=K_{1}=K_{2}$.
We use $\bar{k}_{-1}=10$ in (a, d), $\bar{k}_{-1}=\koff=1$ in (b, c, e), and $\bar{k}_{-1}=\koff=0.1$ in (f), respectively.
We use the same initial values, $\R(t=0)=\Rtot$, $\Rp(t=0)=0$, $\Co{i}(t=0)=0$ (for $i=0, 1, ..., n$), $\Ph(t=0)=\Ptot$, $\D(t=0)=0$, and $\Li(t=0)=\Ltot$.
}\label{fig:zp}
\end{center}
\end{figure*}

\Fig \ref{fig:zp}(a) shows the steady state response of the $\zp$ kinetics (the $\gp$ model with $n=1$). 
The phosphorylated receptor fraction $\Rpstr$ is plotted as a function of unbinding rate $k_{-1}$ for different values of $K$.
Here, we define $K$ as $K:=K_{1}=K_{2}$.
$K$ corresponds to the unsaturation levels of the phosphorylation and dephosphorylation reactions in the $\zp$ kinetics. 
When $K$ is small, the reaction is saturated and becomes zero-order.
When $K$ is large, the reaction is unsaturated and becomes first-order.
As shown in \fig \ref{fig:zp}(a), as $K$ decreases, $\Rpstr(\koff)$ also becomes sigmoidal.
We also plot the phosphorylated receptor fraction at the steady state as a function of the total amount of ligand, $\Rpstr:=\Rpstr(\Ltot)$ in \fig \ref{fig:zp}(b) for various values of $K$. 
As $K$ decreases, $\Rpstr(\Ltot)$ also becomes sigmoidal.
These results indicate that, although the $\mkp$ kinetics has a trade-off between specificity and sensitivity, the $\zp$ kinetics has no trade-off between these properties.

Finally, we examine the speed of the $\zp$ kinetics from time evolution of $\Rpr(t)$ for various values of $K$. 
In \fig \ref{fig:zp}(c), $\Rpr(t)$ is plotted for different values of $K$. 
When we focus on the initial speed, the initial speed at $K=10$ and $K=100$ are faster than that at $K=1$. 
However, the speed at high values of $K$ slows down when approaching the stationary state and becomes slower than that at low values of $K$ until $\Rpr(t)$ reaches the steady state.
In addition, the time for convergence to the steady-state at low values of $K$ is less than that at high values of $K$. 
The steady-state intensity of the response, $\bar \Rpr$, at low values of $K$ is also higher than that at high values of $K$.

From these results, in the $\zp$ kinetics, the decrease in the unsaturation level $K$ amplifies both specificity (\fig \ref{fig:zp}(a)) and sensitivity (\fig \ref{fig:zp}(b)) without the loss of response speed (\fig \ref{fig:zp}(c)) (see also \tab \ref{tab:table1}). 

\subsubsection{Mechanisms of the $\zp$ kinetics to amplify specificity and sensitivity}
The steady state responses of the $\zp$ kinetics show that the $\zp$ kinetics can have both high specificity and sensitivity despite having only a single proofreading step (\figs \ref{fig:zp}(a, b)). 
To understand the mechanism underlying the $\zp$ kinetics, we analyze the reaction fluxes of phosphorylation and dephosphorylation reactions in the $\gp$ model. 

Here, we represent the fluxes of phosphorylation and dephosphorylation reactions in the $\gp$ model, $\F{\gp}$ and $\G {\gp}$ in \eq \eqref{eq:zp-qss-2},  in the forms as
\begin{eqnarray}
&& \F{\gp} = V_{1} \frac{\R}{K_{m1}+\R}, \label{eq:ZP-MM1} \\
&& \G{\gp} = V_{2} \frac{\Rp}{K_{2}+\Rp}, \label{eq:ZP-MM2}
\end{eqnarray}
where $K_{m1}=K_{1}/\sum_{i=0}^{n}\alpha^{i}$.
\Eqs \eqref{eq:ZP-MM1} and \eqref{eq:ZP-MM2} are Michaelis--Menten equations, where $V_{1}$ and $V_{2}$ are the maximum fluxes of $\F{\gp}$ and $\G{\gp}$, respectively. 
$K_{m1}$ and $K_{2}$ are Michaelis--Menten constants. 

The phosphorylation ratio $\Rpstr$ becomes sigmoidal to $V_{1}/V_{2}$ when both the phosphorylation and dephosphorylation reactions  operate in the zero-order region \cite{1981PNAS...78.6840G}.
In the $\gp$ model, $V_{1}/V_{2}$ is represented as follows:
\begin{eqnarray}
\frac{V_{1}}{V_{2}} = \frac{ \kv  \Ltot {\alpha^{n}}/{\sum_{i=0}^{n}\alpha^{i}} }{ k_{3} \Ptot },
\label{eq:GP-V}
\end{eqnarray}
where $\alpha=\kv/(k_{-1}+\kv)$.
In the $\pdpc$ kinetics, the number of proofreading steps is $n=0$, and $V_{1}/V_{2}=\kv  \Ltot / k_{3} \Ptot$, which indicates that $V_{1}/V_{2}$ is dependent on the total amount of ligands $\Ltot$.
Because the $\pdpc$ kinetics becomes ultrasensitive to $V_{1}/V_{2}$ when the phosphorylation and dephosphorylation reactions are in saturating conditions, $K_{m1} \ll \Rtot$ and $K_{2} \ll \Rtot$, $\Rpstr$ of the $\pdpc$ kinetics becomes sigmoidal to $\Ltot$.
On the other hand, in the $\gp$ model with $n \geq 1$, $V_{1}/V_{2}$ depends on both $\Ltot$ and $\koff$ because $\alpha$ depends on $\koff$. As in the case of the $\pdpc$ kinetics, the $\gp$ model, which includes the $\zp$ kinetics, also becomes ultrasensitive, therefore $\Rpstr$ of the $\gp$ model with $n \geq 1$ becomes sigmoidal to both $\Ltot$ and $\koff$.
This is the mechanism by which the $\gp$ model, or the $\zp$ kinetics, has both sensitivity and specificity, whereas the $\pdpc$ kinetics only has sensitivity.

In contrast to the $\pdpc$ kinetics and the $\gp$ model, $\Rpstr$ of the $\mkp$ kinetics is not sigmoidal to $V_{1}/V_{2}$, because the dephosphorylation reaction is always first-order.
Thus $\Rpstr$ of the $\mkp$ kinetics cannot be sigmoidal to $V_{1}/V_{2}$.
In summary, whereas $V_{1}/V_{2}$ of the $\mkp$ kinetics depends on both $\Ltot$ and $\koff$, $\Rpstr$ of the $\mkp$ kinetics   cannot be sigmoidal to $\Ltot$ and $\koff$ due to the first-order characteristics of the dephosphorylation reaction.

\subsubsection{Steady-state response of the ZP kinetics with several proofreading steps}

Next, we investigate how the specificity and sensitivity of the $\zp$ kinetics changes as the proofreading step $n \geq 0$ increases when $K$ is small.
\Fig \ref{fig:zp}(d) shows the steady-state response of the phosphorylated receptor fraction $\Rpstr(\koff)$ as a function of the unbinding rate $k_{-1}$ for $n=0, 1, 2,$ and $3$.
When $n=0$ (red curve), which corresponds to the $\pd$ kinetics, $\Rpstr(k_{-1})$ has shallow slopes. 
In contrast, when $n=1$ (blue curve), which corresponds to the $\zp$ kinetics, the curve of $\Rpstr(k_{-1})$ becomes steep and exhibits an all-or-none switch-like response.
The steepness of $\Rpstr(\koff)$ becomes steeper when $n=2$ than when $n=1$.
However, the increase in $n$ also shifts the threshold of the all-or-none response towards $k_{-1}=0$, and the maximum value of $\Rpstr(k_{-1})$ is decreased when $n\geq4$.
All together, these results indicate that, in the $\zp$ kinetics, there is an optimal number of proofreading steps $n$ for specificity amplification.

We also plot the phosphorylated receptor fraction at the steady state as a function of the total amount of ligand, $\Rpstr:=\Rpstr(\Ltot)$, in \fig \ref{fig:zp}(e) for various values of $n$.
For all values of $n$, the shapes of $\Rpstr(\Ltot)$ are sigmoidal.
However, as $n$ increases, the steepness is decreased.
In addition, the threshold values of the sigmoidal curves are shifted towards larger $\Ltot$ for larger $n$, which means that the system requires more ligands to initiate a response. 
These results indicate that, in the $\zp$ kinetics, the sensitivity monotonically decreases with respect to the number of proofreading steps $n$.

Next, we evaluate the dependence of the speed of the $\zp$ model on $n$ for small $K$. 
In \fig \ref{fig:zp}(f), $\Rpr(t)$ is plotted for various values of $n$ obtained from the numerical simulation of \eqs \eqref{eq:ZP-1} -- \eqref{eq:ZP-6}.
The transition time to the stationary state of $\Rpr(t)$ is monotonically delayed with the increase in $n$. 
This result indicates that the increase in the proofreading step $n$ leads to a loss of speed in the $\zp$ kinetics.

Taken together, we find that there is an optimal $n$ to amplify the specificity of the $\zp$ kinetics (\fig \ref{fig:zp}(d)). 
In addition, the $\zp$ kinetics can amplify specificity by introducing an additional proofreading step $n$ at the cost of decreasing the sensitivity and speed.
Therefore, the $\zp$ kinetics has an optimal $n$ that balances specificity, sensitivity, and speed.

\subsection{Quantitative analysis of the specificity, sensitivity, and speed of the $\gp$ model}
We have shown that the unsaturation levels of phosphorylation and dephosphorylation and the number of proofreading steps $n$ affect the specificity, sensitivity, and speed. 
In particular, the saturating condition enables the $\zp$ kinetics to increase specificity even though the system has only a single proofreading step.
However, we do not yet know an optimal combination of saturation and the number of proofreading steps $n$ to balance high levels of discrimination efficiencies. 
To understand the optimal strategy of proofreading, we quantitatively evaluate specificity, sensitivity, and speed as functions of both the unsaturation level $K$ and the number of proofreading steps $n$. 

First, we define the specificity function $\lambda(n, K)$ as 
\begin{eqnarray}
\lambda(n, K):=\left. 
	{\Rpstr}_{\rm max} \frac{\diff \Rpstr(k_{-1})}{\diff k_{-1}} 
	\right|_{\Rpstr = {\Rpstr}_{\rm max}/2},
	\label{eq:specificity}
\end{eqnarray}
where ${\Rpstr}_{\rm max}:={\rm max} {\Rpstr(k_{-1})}$.
Next, we define the sensitivity function $\mu(n, K)$ as
\begin{eqnarray}
\mu(n, K):=\left. 
	{\Rpstr}_{\rm max} \frac{\diff \Rpstr(\Ltot)}{\diff \Ltot}  
	\right|_{\Rpstr = {\Rpstr}_{\rm max}/2},
\label{eq:sensitivity}
\end{eqnarray}
where ${\Rpstr}_{\rm max}:={\rm max} {\Rpstr(\Ltot)}$. 
$\lambda(n, K)$ and $\mu(n, K)$ are the products of the maximal response intensity, ${\Rpstr}_{\rm max}$, and the steepness of $\Rpstr$. 
Therefore, even if the slope of $\Rpstr$ is steep, $\lambda(n, K)$ and $\mu(n, K)$ can be small when ${\Rpstr}_{\rm max}$ is small.
We also define the speed function $\nu(n, K)$ as 
\begin{eqnarray}
\nu(n, K):=\frac{{\Rpstr}_{\rm 0.9}}{t_{0.9}}, \label{eq:speed}
\end{eqnarray}
where ${\Rpstr}_{\rm 0.9}:=0.9\, {\rm max} {\Rpstr}$.
Here, $t_{0.9}$ is the first passage time to reach $90\%$ of the maximum value of $\Rpr$ and therefore it satisfies $\Rpr(t_{0.9})={\Rpstr}_{\rm 0.9}$.
Using these quantitative measures of specificity, sensitivity, and speed, we analyze the optimal parameter region for ligand discrimination in the $\gp$ model.

\begin{figure*}[!htbp]
\begin{center}
          \includegraphics[width=\textwidth]{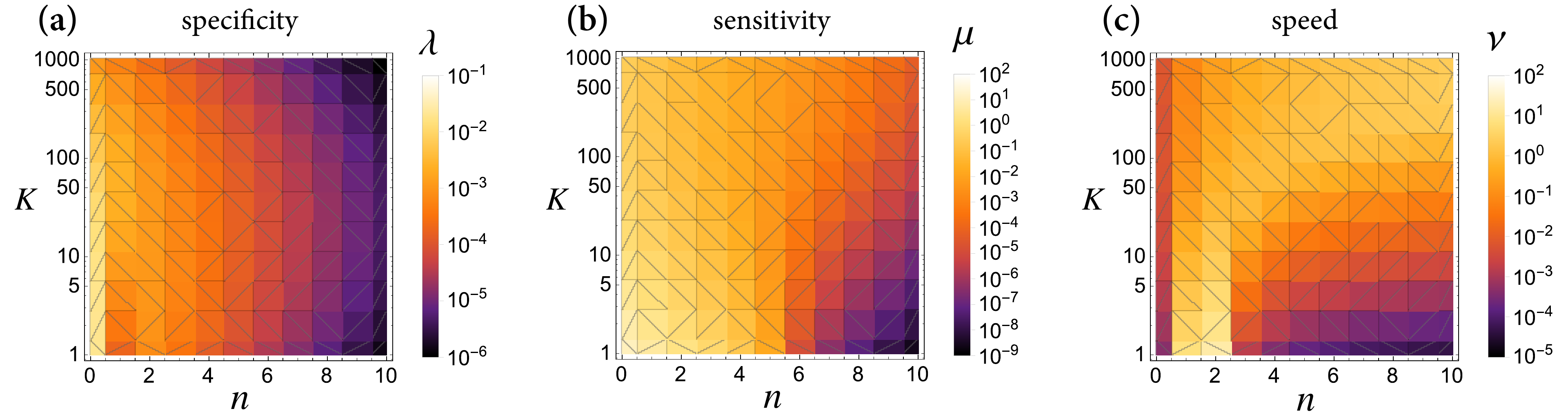}
          \hspace{1.6cm} 
\caption{
The specificity, sensitivity, and speed of the $\gp$ model as functions of the number of proofreading steps $n$ and the unsaturation level $K$ are plotted as color maps.
(a) 
The specificity function $\lambda(n, K)$ with $\Ltot=3$. 
(b) 
The sensitivity function $\mu(n, K)$ with $\koff=1$. 
(c)
The speed function $\nu(n, K)$ with $\Ltot=3$ and $\koff=1$. 
In (a) and (b), $\lambda(n, K)$ (a) and $\mu(n, K)$ (b) are calculated from $\Rpstr(k_{-1})$. 
$\Rpstr(k_{-1})$ and $\nu(n, K)$ (c) are obtained from the numerical simulations of \eqs \eqref{eq:ZP-1} -- \eqref{eq:ZP-7}.
We use the initial states indicated in \fig \ref{fig:zp}.
The other parameters are $\Rtot=100$, $\Ptot=1$, $\kv=1$, $k_{-2}=10$, and $k_{3}=1$.
In (a) -- (c), $k_{1}$ and $k_{2}$ are obtained as $k_{1}=(\bar{k}_{-1}+\kv)/K_{1}$ and $k_{2}=(k_{-2}+k_{3})/K_{2}$ where $K=K_{1}=K_{2}$ and $\bar{k}_{-1}=10$.
}
\label{fig:3chara}
\end{center}
\end{figure*}

\Fig \ref{fig:3chara}(a) plots the specificity function $\lambda(n, K)$ of the $\gp$ model. 
When $K$ is large, which corresponds to first-order reactions, the optimal number of proofreading steps $n$ that maximizes $\lambda(n, K)$, is high. 
For small values of $K$, which correspond to zero-order reactions, the optimal $n$ shifts towards to $n=1$. 
In addition, $\lambda(n, K)$ is maximized when $K$ is small and $n=1$ and $n=2$. 
Therefore, in the $\gp$ model, the specificity $\lambda(n, K)$ can be maximized when both $n$ and $K$ are small.

\Fig \ref{fig:3chara}(b) shows the sensitivity function $\mu(n, K)$ of the $\gp$ model. 
For all values of $K$, $\mu(n, K)$ monotonically decreases with increases in $n$.
Thus $\mu(n, K)$ is maximized when $n=0$ for all $K$.
When we fix $n=0$, $\mu(K):=\mu(0,K)$ is maximized when $K$ is small ($K=1$).
Therefore, \fig \ref{fig:3chara}(b) indicates that the sensitivity $\mu(n, K)$ is maximized for $n=0$ and small values of $K$, which correspond to the $\pd$ kinetics in the zero-order regime. 

We also plot the speed $\nu(n, K)$ of the $\gp$ model in \fig \ref{fig:3chara}(c). 
Generally, $\nu(n, K)$ with fixed $K$, $\nu(n):=\nu(n, K)$, decreases with increases in $n$ for all $K$.
$\nu(n, K)$ monotonically increases as $K$ decreases when $n=0$.
On the other hand, $\nu(n, K)$ with fixed $n\geq1$, $\nu(K):=\nu(n, K)$ for $n\geq1$, has an optimal $K$ that maximizes $\nu(K)$. 
However, the curves of $\nu(K)$ for $n\geq1$ are shallow.
Therefore, $\nu(n, K)$ is maximized when $n=0$ and $K$ is small, meaning that the speed is maximized when the system has few proofreading steps and operates in the zero-order regime.

\begin{figure}[!hbtp]
\begin{center}
\includegraphics[width=0.45\textwidth]{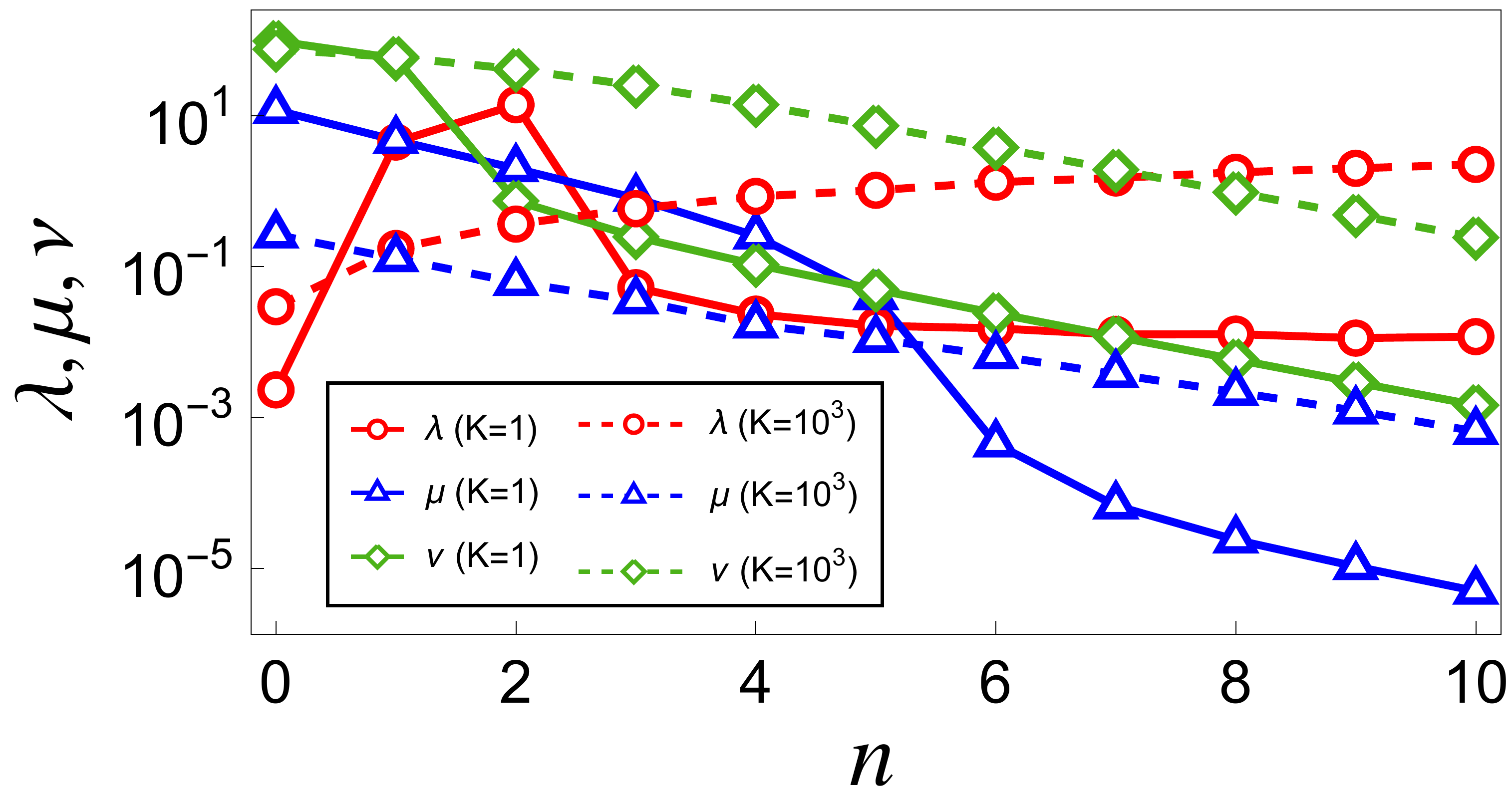}
\caption[]{
The specificity $\lambda(n,K)$ (blue), sensitivity $\mu(n,K)$ (orange), and speed $\nu(n,K)$ (green) as functions of the number of proofreading steps $n$ for different unsaturation levels, $K=1$ (solid lines) and $K=1000$ (dashed lines), which correspond to the zero-order regime and the first order regime, respectively.
The data plotted are the same as in \fig \ref{fig:3chara}.}
\label{fig:3CharasVSn_GP.pdf}
\end{center}
\end{figure}

To find the optimal parameter region to balance the three discrimination properties, we plot   specificity $\lambda(n, K)$ (red curves), sensitivity $\mu(n, K)$ (blue curves), and speed $\nu(n, K)$ (green curves) as functions of $n$ for both $K=1$ and  $K=1000$ as in \fig \ref{fig:3CharasVSn_GP.pdf}. 
The data plotted in \fig \ref{fig:3CharasVSn_GP.pdf} are selected from the data of $K=1$ and $K=1000$ used in \fig \ref{fig:3chara}.
Small $K$ ($K=1$, solid curves) and large $K$ ($K=1000$, dashed curves) correspond to the zero-order reaction and the first-order reaction, respectively. 
$\lambda(n, K)$ with $K=1$ is maximized when $n=2$ and the maximum value is much larger than that of $\lambda(n, K)$ with $K=1000$ and $n=10$. 
For both small and large $K$, $\mu(n, K)$ monotonically decreases with increasing $n$, and $\mu(n, K)$ with $K=1$ is much larger than that with $K=1000$ for $0 \leq n \leq 5$. 
Similarly to $\mu(n, K)$, $\nu(n, K)$ also monotonically decreases with $n$. 
On the other hand, $\nu(n, K)$ with $K=1$ has almost the same value as that with $K=1000$ for $0 \leq n \leq 1$, and $\nu(n, K)$ with $K=1$ is much lower than that with $K=1000$ for $n \geq 1$. 
Therefore, when $n=1$, the system with small $K$ has much higher specificity and sensitivity than that with large $K$, and the speed with small $K$ is at the same level as that with large $K$. 
Thus, small $K$ and $n=1$, and $2$ represent the optimal parameter region of the $\gp$ model to balance the specificity, sensitivity and speed at high levels.

\section{Concentration compensation for ligand discrimination}

We have investigated ligand discrimination systems for a fixed unbinding rate $k_{-1}$ and total ligand concentration $\Ltot$. 
These conditions correspond to the following situations: a cell discriminates between the target and non-target ligands when both concentrations are the same ($\Ltot$ is fixed); a cell detects a low concentration of its target ligand ($k_{-1}$ is fixed).
However, a more realistic situation is that a cell selectively detects a low concentration of the target ligands in the presence of a high concentration of non-target ligands. 
Therefore, we need to evaluate discrimination efficiencies under the condition that both the unbinding rate $k_{-1}$ and the total amount of ligand $\Ltot$ are not fixed.
A ligand discrimination system should be activated by a low amount of its target ligands and should not   be activated by a large amount of non-target ligands. 
Such properties can be realized at least if the threshold of $\koff$ is independent of $\Ltot$.
For simplicity, we consider the situation that the number of ligand types in a reaction system is $1$.
In this situation, we consider the problem where a cell needs to discriminate whether a given ligand is a  target or non-target ligand.

\begin{figure*}[!htbp]
\begin{center}
\includegraphics[width=\textwidth]{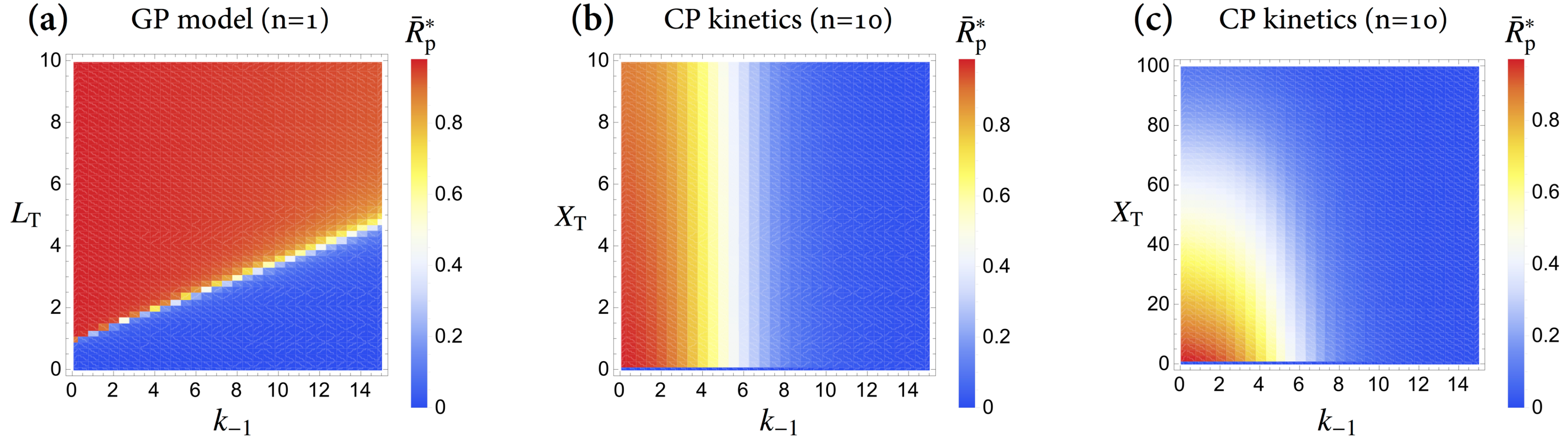}
\end{center}
\caption{
The steady-state fraction of the phosphorylated receptor $\Rpstr$ of the $\gp$ model with $n=1$ (the $\zp$ kinetics) (a) and the $\cp$ kinetics with $n=10$ (b, c) as a function of the unbinding rate $k_{-1}$ and the total ligand concentration $\Ltot$ (a) or $\Xtot$ (b, c). 
$\Rpstr$ is shown by colors, and is obtained from the numerical simulation of \eqs \eqref{eq:ZP-1} -- \eqref{eq:ZP-7} (a) and \eqs \eqref{eq:cp1} -- \eqref{eq:cp7} (b, c).
The initial states are as in \fig \ref{fig:zp} (a) and \fig \ref{fig:CP3quaChara} (b, c).
In (c), the data are the same as those of (b), but the plotted region is changed from $0\leq \Xtot \leq 10$ to $0\leq \Xtot \leq 100$.
The parameters are $\Rtot=100$, $k_{1}=10$, $k_{2}=10$, and $k_{-2}=10$.
The other parameters are $\Ptot=1$, $\kv=2$, and $k_{3}=1$ (a), and $\kv=10$ and $k_{3}=0.1$ (b, c).
}
\label{fig:ALD_CP}
\end{figure*}

\Fig \ref{fig:ALD_CP}(a) shows the steady-state phosphorylated receptor ratio $\Rpstr$ of the $\gp$ model with $n=1$ (the $\zp$ kinetics), as a function of both the unbinding rate $k_{-1}$ and the ligand concentration $\Ltot$ in the zero-order regime. 
$\Rpstr$ is small for small $k_{-1}$ and small $\Ltot$.
However, $\Rpstr$ becomes large for large $k_{-1}$ and large $\Ltot$ because the threshold of the all-or-none response of $\Rpstr$ is shifted by $\Ltot$. 
Thus, the $\gp$ model can respond to a high concentration of a non-target ligand. 
Therefore, the $\gp$ model can not discriminate a low amount of the target ligand from a large amount of non-target ligands. 

This property can be understood by analyzing the fluxes of phosphorylation and dephosphorylation reactions. 
As we discussed above, the $\gp$ model has sensitivity to $V_{1}/V_{2}$.
As shown in \eq \eqref{eq:GP-V}, $V_{1}/V_{2}$ of the $\gp$ model depends on both $\koff$ and $\Ltot$, because $\alpha$ depends on $\koff$.
Therefore, when the phosphorylation and dephosphorylation reactions operate in the zero-order region, the steady-state phosphorylated receptor ratio $\Rpstr$ can be sensitive to both $\koff$ and $\Ltot$. 
Here, $V_{1}/V_{2}$ monotonically decreases with decreasing $k_{-1}$, whereas it monotonically increases with increasing $\Ltot$. 
Then, even though $k_{-1}$ is large, which corresponds to a non-target ligand, the decrease of $V_{1}/V_{2}$ can be compensated for by the increase of $\Ltot$. 
This is the mechanism by which the $\gp$ model responds to a high concentration of non-target ligands.

\subsection{Kinetic proofreading with ligand concentration compensation}

To realize insensitivity to a large amount of non-target ligands, we derive another model from the $\zp$ kinetics.
For convenience, we denote the model as the ligand concentration compensation proofreading model (the $\cp$ kinetics).
By introducing an assumption that ligand $\rm \Li$ and phosphatase $\rm \Ph$ are regarded as the same molecule $\rm X$, we derive the $\cp$ kinetics from the $\gp$ model (the $\zp$ kinetics with $n\geq0$) as follows:
\begin{equation}\label{model:CP}
\includegraphics[width=0.47\textwidth]{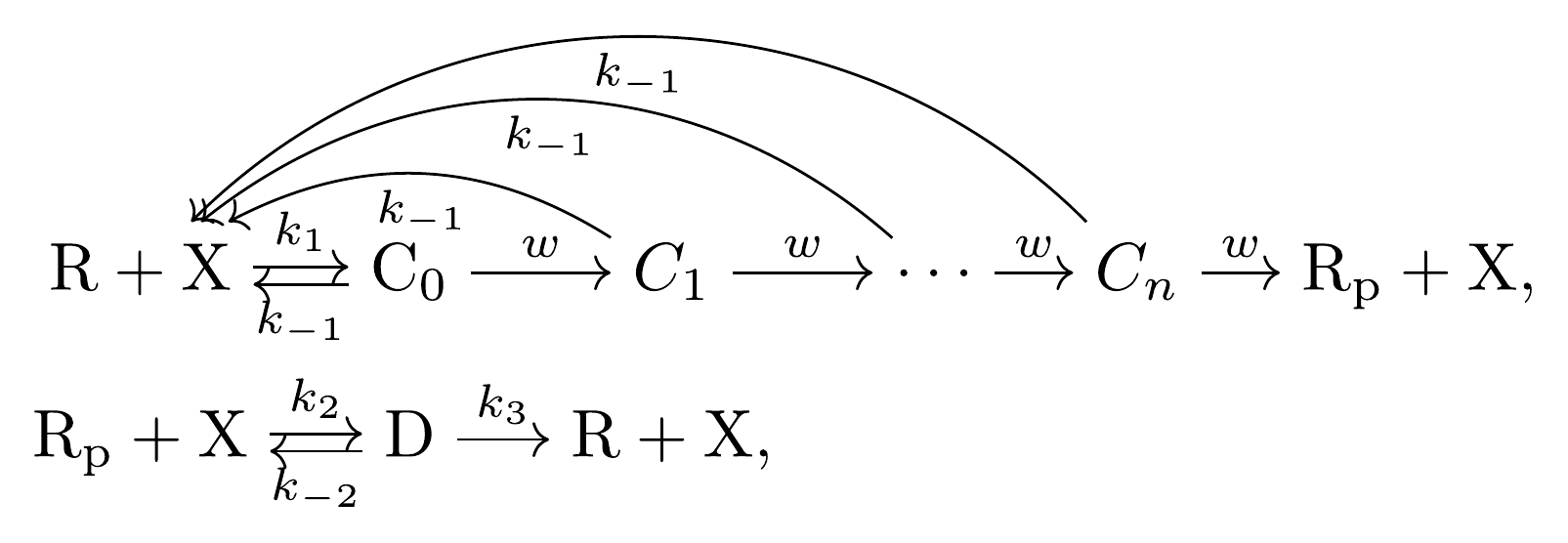}
\end{equation}
where $\rm \X$ is a molecule that works as both a ligand and the enzyme to induce or catalyze receptor phosphorylation and dephosphorylation. $\rm \Co{i}$ (for $i=0, \dots, n$) are complexes of $\rm \R$ and $\rm \X$, and $\rm \D$ is a complex of $\rm \Rp$ and $\rm \X$. 

The dynamics of $\cp$ kinetics can be represented by the following ODEs:
\begin{align}
 \ddf \R & = - k_{1} \R \X + k_{3} \D + k_{-1} \sum_{i=0}^{n}{\Co{i}},\label{eq:cp1}\\
 \ddf \X &= - k_{1} \R \X + \kv {\Co{n}} + k_{-1} \sum_{i=0}^{n}{\Co{i}} - k_{2} \Rp \X \notag \\ &\quad + k_{-2} \D + k_{3} \D, \label{eq:cp2}\\
 \ddf {\Rp} &= \kv {\Co{n}} - k_{2} \Rp \X + k_{-2} \D,\label{eq:cp3}\\
 \ddf \D &= k_{2} \Rp \X - k_{-2} \D - k_{3} \D,\label{eq:cp5}\\
 \ddf {\Co{0}} &= -k_{-1} {\Co{0}} - \kv {\Co{0}} + k_{1} \R \X,\label{eq:cp6}\\
 \ddf {\Co{i}} &= -k_{-1} {\Co{i}}  - \kv {\Co{i}} + \kv {\Co{i-1}}, \, (i=1, ..., n), \label{eq:cp7}
\end{align}
where $\X$ and $\D$ represent the concentrations of the molecules ${\rm \X}$ and ${\rm \D}$, respectively.

From \eqs \eqref{eq:cp1} -- \eqref{eq:cp7}, we derive a Michaelis--Menten approximation of the $\cp$ kinetics. 
If we assume the quasi-steady-state condition such that $\diff \Co{i}/\diff t$ = 0 (for $i=0, 1, \dots, n$) and $\diff \D/\diff t=0$, the dynamics of the $\cp$ kinetics can be represented as follows:
\begin{eqnarray}
&&\frac{\diff \R}{\diff t}
 =  \G{\cp} - \F{\cp},\label{eq:cp-qss2}\\
&&\frac{\diff \Rp}{\diff t}
 =  \F{\cp} - \G{\cp}, \label{eq:cp-qss1}
\end{eqnarray}
where 
\begin{eqnarray}
\F{\cp}:= \frac{\kv \alpha^{n} K_{2} \Xtot \R}{K_{1} K_{2} + K_{2} \sum_{i=0}^{n} \alpha^{i} \R + K_{1} \Rp},\\
\G{\cp}:=\frac{k_{3} K_{1} \Xtot \Rp}{K_{1} K_{2} + K_{2} \sum_{i=0}^{n} \alpha^{i} \R + K_{1} \Rp}.
\end{eqnarray}
(see also \tab \ref{tab:table1}). 
Here $\Xtot:=\X + \sum_{i=0}^{n}\Co{i}+\D$ is the total ligand concentration.

When $\cp$ kinetics is in the steady-state, that is $\diff \R/\diff t =0$ and $\diff \Rp/\diff t =0$, the steady-state phosphorylated receptor fraction $\Rpstr$ can be described as
\begin{eqnarray}\label{eq:cp-sss}
\Rpstr = \frac{1}{1+  k_{3} K_{1}/\kv \alpha^{n} K_{2}}.
\end{eqnarray}
Here we assume that the total receptor concentration $\Rtot$ is much larger than the total ligand concentration $\X$, that is, $\Rtot \gg \Xtot$ where $\Rtot:=\R+\Rp+\sum_{i=0}^{n}\Co{i}$. 
This result clearly indicates that the steady-state phosphorylated receptor ratio $\Rp$ of the $\cp$ kinetics does not depend on the total ligand concentration $\Xtot$ but depends on the unbinding rate $k_{-1}$ because $\alpha$ is a function of $\koff$. 
This approximation suggests that the $\cp$ kinetics can be a model for insensitivity to non-target ligands.

\begin{figure*}[!htbp]
\begin{center}
          \includegraphics[width=1\textwidth]{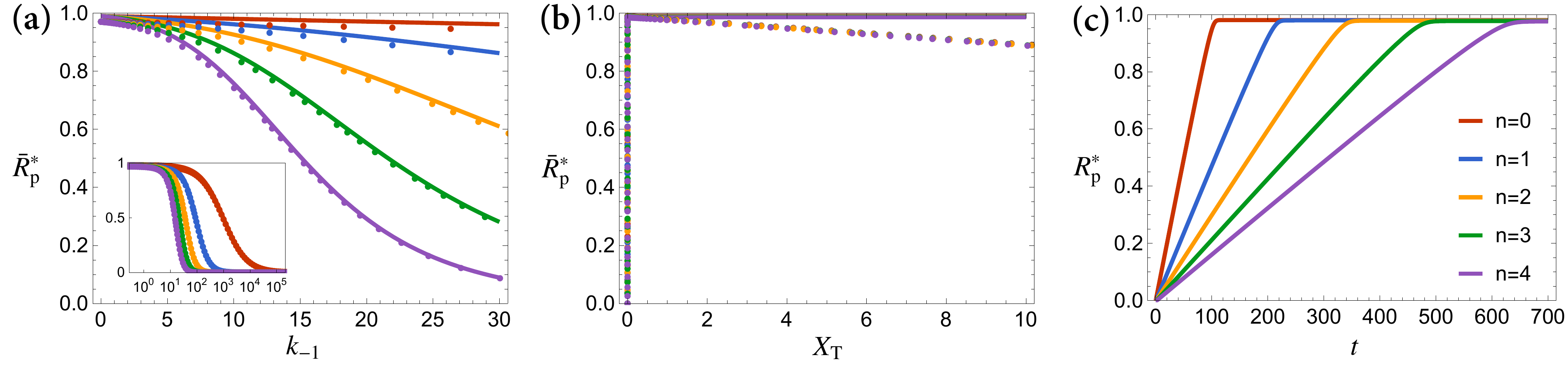}
\caption{
The steady-state response and the time series of the $\cp$ kinetics for various numbers of proofreading steps $n$: $n=0$ (red), $n=1$ (blue), $n=2$ (yellow), $n=3$ (green), and $n=4$ (purple).
(a, b) 
The stationary fraction of the phosphorylated receptor $\Rpstr$ as a function of the unbinding rate $\koff$ with $\Xtot=2$, $k_{1}=10$, $k_{2}=10.1$, and $k_{3}=0.1$ (a), and $\Rpstr$ as a function of the total ligand concentration $\Ltot$ with $k_{1}=11$, $\koff=1$, $k_{2}=10.1$, and $k_{3}=0.1$ (b).   
The solid curves are analytically obtained from \eq \eqref{eq:cp-sss} and the dotted curves are obtained from the numerical simulation. 
The semilogarithmic plot is also shown in the inset of (a).
(c) 
The time series of the fraction of phosphorylated receptor $\Rpr(t)$ with $\Ltot=5$, $k_{1}=10$, $\koff=0.1$, $k_{2}=1$, and $k_{3}=1$. 
In (a) -- (c), the values of the other parameters are $\Rtot=100$, $k_{1}=10$, $\kv=1$, and $k_{-2}=10$.
The results of the numerical simulations are obtained from \eqs \eqref{eq:cp1} -- \eqref{eq:cp7}. 
For the numerical simulations, we use the same initial values, $\R(t=0)=\Rtot$, $\Rp(t=0)=0$, $\Co{i}(t=0)=0$ (for $i=0, 1, \cdots, n$), $\D(t=0)=0$, and $\X(t=0)=\Xtot$.
}
\label{fig:CP3quaChara}
\end{center}
\end{figure*}

To verify the specificity and sensitivity of the $\cp$ kinetics, we plot $\Rpstr(\koff)$ as shown in \fig \ref{fig:CP3quaChara}(a, b) for the number of proofreading steps $n$. 
As $n$ increases, $\Rpstr(\koff)$ becomes a highly nonlinear sigmoidal function. 
Therefore, the $\cp$ kinetics with large $n$ has high specificity. 

We also plot $\Rpstr(\Xtot)$ as shown in \fig \ref{fig:CP3quaChara}(b) for $n$. 
For all $n$, $\Rpstr$ can be highly sensitive to $\Xtot$, and the threshold is $\Ltot \approx 0$. 
In addition, the value  of $\Rpstr$ for each $n$ is almost constant for $\Xtot \ll \Rtot$. 
However, the numerical solutions of $\Rpstr$ gradually deviate from the analytical solutions (\eqs \eqref{eq:cp-sss}) with increasing $\Xtot$ because the assumption $\Rtot \gg \Xtot$ is violated. 
As a result, the $\cp$ kinetics has sensitivity to target and also has ligand concentration compensation insensitivity to non-target when the total ligand concentration $\Xtot$ is much lower than the total receptor concentration $\Rtot$. 

In addition, we plot $\Rpr(t)$ as shown in \fig \ref{fig:CP3quaChara}(c) for $n$.
The transition time to the stationary state of $\Rpr(t)$ is delayed increasing $n$. 
This result indicates that increases in the number of proofreading steps $n$ lead to losses of speed in the $\cp$ model.
From these results, although we find that the specificity of the $\cp$ kinetics is increased and the sensitivity is not reduced with increasing $n$, there is a trade-off between the specificity and the speed with increasing $n$.

Next, we plot $\Rpstr(\koff, \Xtot)$ of the $\cp$ kinetics with $n=10$ as shown in \fig \ref{fig:ALD_CP}(b, c). 
While $\Rpstr$ of the $\gp$ model shown in \fig \ref{fig:ALD_CP}(a) depends on both $\koff$ and $\Ltot$, $\Rpstr$ of the $\cp$ kinetics depends only on $k_{-1}$ when $\Xtot>0$. 
In addition, $\Rpstr$ of the $\cp$ kinetics is $\Rpstr=0$ when $\Xtot=0$ for all $k_{-1}$. 
This data show that the $\cp$ kinetics with large values of $n$ has not only high specificity but also high sensitivity to target ligands and low sensitivity to non-target ligands.
Therefore, from \figs \ref{fig:ALD_CP}(b) and \ref{fig:CP3quaChara}, although the speed is slow, the $\cp$ kinetics with large values $n$ has the specificity, sensitivity to target and insensitivity to non-target.
However, as we show in \fig \ref{fig:ALD_CP}(b), the concentration compensation does not hold when $\Xtot \approx \Rtot$. 
Generally, this condition may not appear in T cells because the total number of TCRs is much larger than that of ligands.
In summary, the $\cp$ kinetics with large values of $n$ can be a\delkj{n absolute ligand discrimination} model \delkj{with}\addkj{for} high specificity, sensitivity \addkj{to target} and \delkj{ligand concentration compensation} \addkj{insensitivity to non-target} when $\Xtot \ll \Rtot$.

%% file: discussion.tex
\section{Summary \& Discussion}
The important characteristics of a ligand discrimination system are specificity, sensitivity to target, quick discrimination, and insensitivity to non-target. 
The multistep kinetic proofreading scheme only has specificity, and therefore other mechanisms that can balance all of the four characteristics should be considered.
Altan-Bonet proposed a kinetic proofreading model with feedback regulation to balance both specificity and sensitivity \cite{Altanbonnet:2005hoa, 2013PNAS..110E.888F}, then Fran{\c c}ois and colleagues  performed \textit{in silico} evolution to obtain an ``adaptive sorting'' mechanism for specificity, sensitivity to target, and insensitivity to non-target \cite{2016JSP...162.1130F,2013PhRvL.110u8102L}.
These studies indicate that all or some of the four discrimination properties are generated by nonlinear responses of biochemical reaction systems.
However, other possible mechanisms that could give rise to the four characteristics have not yet been fully investigated.

In this study, we demonstrated that zero-order ultraspecificity, which is derived from the zero-order ultrasensitivity, can be an alternative mechanism to amplify specificity.
By using the zero-order ultraspecificity mechanism, the $\zp$ kinetics can have higher specificity than the $\mkp$ kinetics, even though the $\zp$ kinetics has only a single proofreading step whereas the $\mkp$ kinetics has multiple proofreading steps.
Due to the zero-order ultrasensitivity in the ultraspecificity mechanism, the $\zp$ kinetics also has sensitivity to ligand concentration. 
In addition, as the zero-order ultraspecificity mechanism does not require multiple proofreading steps to enhance specificity and sensitivity, the mechanism is also advantageous for response speed.

We also investigated the optimal conditions for ligand discrimination using the $\gp$ model, which is a generalized form of the $\mkp$ kinetics and the $\zp$ kinetics. 
The results indicate that the optimal condition that balances specificity, sensitivity, and speed can be realized when the system is in a zero-order regime and the number of proofreading steps is small but non-zero.

The $\zp$ kinetics has specificity, sensitivity, and speed. 
However, the $\zp$ kinetics does not have insensitivity to non-target, which is essential for unresponsiveness to a high concentration of non-target ligands. 
To understand the mechanisms, we showed that the $\cp$ kinetics for insensitivity to non-target is naturally derived from the $\zp$ kinetics by assuming that the ligand and phosphatase can be regarded as the same molecule. 
The $\cp$ kinetics requires several proofreading steps to amplify specificity and has a trade-off between specificity and speed. 
However, the $\cp$ kinetics with the optimal number of proofreading steps balances all of the four properties, which are required for T cell ligand discrimination.
Therefore, the $\cp$ kinetics can be a possible mechanism for T cell ligand discrimination.

Finally, we summarize the relationship among the five kinetics or models, the approximated functions of phosphorylation-dephosphorylation reactions, and the four characteristics of a ligand discrimination system in \tab \ref{tab:table1}. 
\begin{table*}
\caption{\label{tab:table1}$\F{m}$ and $\G{m}$ for $m \in \{ \rm{\mkp, \pd, \gp, \cp}\}$.}
\begin{ruledtabular}
\begin{tabular}{ccccccc}
 m &\mkp&\pd&\gp\footnotemark[1] &\cp\\ \hline
 $\F{m}$
&$ \frac{\kv \Ltot \alpha^{n}}{\sum_{i=0}^{n}\alpha^{i}} \frac{\R}{(K_{1} / \sum_{i=0}^{n}\alpha^{i}) + \R}$
 &$\kv \Ltot  \frac{\R}{K_{1}+\R}$ 
 &$ \frac{\kv \Ltot \alpha^{n}}{\sum_{i=0}^{n}\alpha^{i}} \frac{\R}{(K_{1}/\sum_{i=0}^{n}\alpha^{i})+\R}$
 &$ \frac{\kv \alpha^{n} K_{2} \Xtot \R}{K_{1} K_{2} + K_{2} \sum_{i=0}^{n} \alpha^{i} \R + K_{1} \Rp} $ 
 \\
 $\G{m}$ 
 &$k_{4} \Rp$
 &$k_{3} \Ptot \frac{\Rp}{K_{2}+\Rp}$
 &$k_{3} \Ptot \frac{\Rp}{K_{2}+\Rp}$
 &$\frac{k_{3} K_{1} \Xtot \Rp}{K_{1} K_{2} + K_{2} \sum_{i=0}^{n} \alpha^{i} \R + K_{1} \Rp}$
 \\ \hline
 specificity   
 & $\bigcirc$
 & $\times$
 & {\large $\circledcirc$}\footnotemark[2]
 & $\bigcirc$
 \\
 sensitivity   
 & $\times$
 & {\large $\circledcirc$}
 & $\bigcirc$
 & $\bigcirc$
 \\
 speed
 & $\times$
 & $\bigcirc$
 & $\bigcirc$\footnotemark[2]
 & $\bigtriangleup$\footnotemark[3]
 \\
 insensitivity\footnotemark[4]
 & $\times$
 & $\times$
 & $\times$
 & $\bigcirc$
\end{tabular}
\end{ruledtabular}
\footnotetext[1]{The $\gp$ model includes the $\zp$ kinetics in saturating conditions.}
\footnotetext[2]{the $\zp$ kinetics (for small $n$ ($n \neq 0$))}
\footnotetext[3]{the speed is decreased with $n$}
\footnotetext[4]{insensitivity means insensitivity to non-target ligand concentration}
\end{table*}

\subsection{Biological relevance of the $\zp$ kinetics and the $\cp$ kinetics}
Although the $\zp$ kinetics and the $\cp$ kinetics are simple and seem to be biologically realistic enough, experimental verification is necessary to determine whether these kinetics are actually used in real ligand discrimination systems. 
Here we discuss the biological relevance by focusing on T cell ligand discrimination.

Both the $\zp$ kinetics and the $\cp$ kinetics are based on a receptor phosphorylation and dephosphorylation cycle. 
TCRs form a complex with CD3 molecules, which have several modification sites in immunoreceptor tyrosine-based activation motifs. 
The TCR activation and inactivation can be summarized as follows: The modification sites are phosphorylated by the Src family tyrosine kinases Lck. 
On the other hand, the receptor tyrosine phosphatase CD45 dephosphorylates the TCR modification sites \cite{VanderMerwe2011}. 
These findings from molecular biology support that TCRs have a phosphorylation and dephosphorylation cycle.

In the $\zp$ kinetics, we do not clearly model the activation enzymes such as Lck.
Instead, we assume that ligand $\rm \Li$ itself catalyzes the receptor activation reaction. 
Similarly to TCRs, the co-receptors CD4 and CD8 are also embedded in the plasma membrane \cite{Anton-Omer:2012dd}. 
The co-receptors is considered to bind to peptide-major histocompatibility complex (pMHC) \cite{Li2004, Anton-Omer:2012dd}, which is a complex of a ligand and MHC on the surface of an antigen-presenting cell, and Lck is considered to associate with the co-receptors \cite{2010PNAS..10716916A, Li2004, Anton-Omer:2012dd}. 
Therefore, the assumption that ligand $\rm \Li$ itself has a catalytic function is reasonable if we admit that the complex of pMHC, CD4 or CD8, and Lck is necessary for TCR phosphorylation.

It should be noted that the data are not yet available to say with certainty that the ligand-kinase complex and the phosphatase are saturated, which is a necessary condition for the zero-order ultrasensitivity. 
However, T cells can be activated by only about 10 specific ligands \cite{Huang:2013jn,2002Natur.419..845I} and CD45 is excluded from the region where TCR and peptide-MHC are bound after the initiation of contact between a T cell and an antigen-presenting cell \cite{VanderMerwe2011, 2012Natur.487...64J}. 
These facts suggest that the effective concentrations of Lck and CD45, which are a kinase and a phosphatase, respectively, may be much lower than that of TCR. 
Therefore, there is a possibility that the phosphorylation and dephosphorylation may operate in the zero-order regime.

\begin{figure}[!htbp]
\centering
\includegraphics[width=0.45\textwidth]{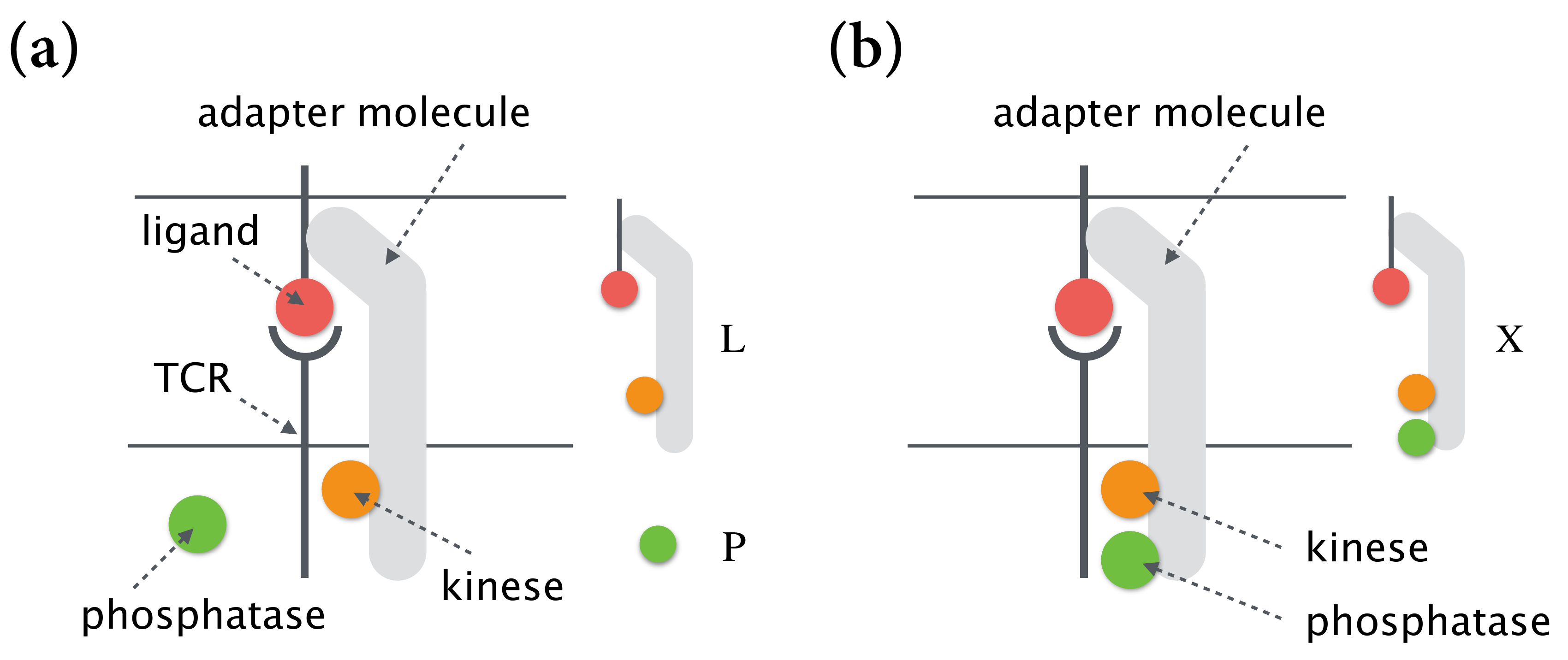}
\caption{Schematic diagram of the $\zp$ kinetics and the $\cp$ kinetics. 
(a) In the $\zp$ kinetics, ligand $\rm L$ is interpreted as a complex of a ligand (pMHC), an adaptor molecule (CD4 or CD8), and a kinase (Lck). 
(b) In the $\cp$ kinetics, molecule $\rm X$ is interpreted as a  complex of a ligand (pMHC), an adaptor molecule (CD4 or CD8), a kinase (Lck), and a phosphatase (CD45).}
\label{fig:compensation.pdf}
\end{figure}

In the $\cp$ kinetics, in contrast, we further assume that the ligand and phosphatase are the same molecule. 
This assumption can be reasonable if we admit that the ligand-coreceptor-kinase complex also associates with a phosphatase and forms a complex as shown in \fig \ref{fig:compensation.pdf}. 
Even though such a complex has not yet been reported, Lck has several modification states and the modifications are regulated by CD45 \cite{Chakraborty2014hw}. 
This fact suggests the possibility to form a complex of pMHC, CD4 or CD8, Lck, and CD45. 
If so, the $\cp$ kinetics can be a possible mechanism for insensitivity to non-target ligands.

\subsection{Possible extensions on the mechanisms of ligand discrimination}

Although we proposed new mechanisms for specificity amplification and insensitivity to non-target ligands, we have not dealt with the stochastic characteristics of a ligand discrimination system. 
The fact that a T cell can be activated by about 10 ligands \cite{Huang:2013jn,2002Natur.419..845I} suggests that there would be a considerable effect of stochastic noise on the discrimination accuracy. 
Recent theoretical research on cellular decision-making in noisy environments connects optimal information processing dynamics and its biological kinetic motif. 
For the detection of a change in the concentration of ligand in a noisy environment, an autocatalytic phosphorylation and autocatalytic dephosphorylation cycle is an optimal kinetics according to information theory, while the PdPc kinetics is suboptimal \cite{2010PhRvL.104v8104K, Kobayashi2011dh}. 
Here, the noisy environment means that the ligand concentration is so low that the number of ligand-receptor complexes fluctuates.
The $\zp$ kinetics, which we propose for ligand discrimination, is a modified form of the $\pdpc$ kinetics.
Thus, the $\zp$ kinetics may not be the best kinetics in an ideal environment, but it may be the better kinetics that is more robust in a noisy environment.
If the $\zp$ kinetics is not so robust to the noisy signal, additional feedback regulation such as the autocatalytic phosphorylation and autocatalytic dephosphorylation may improve the robustness of the $\zp$ kinetics.
In addition, some kinetics with a bifunctional molecule, such as molecule \X\, in the $\cp$ kinetics, are also robust in a noisy environment. 
For example, Mora showed that a push-pull reaction kinetics with multiple kinetic proofreading can robustly sense the target ligand concentration \cite{2015PhRvL.115c8102M}.
This kinetics has a molecule that has two states, active and inactive.
The activation and inactivation reactions form push-pull reactions that are regulated by two different enzymes. 
These enzymes are activated by the same receptor molecules, which have multiple proofreading steps.
This kinetics in \Ref\cite{2015PhRvL.115c8102M} and the $\cp$ kinetics have the similarity that a bifunctional molecule positively and negatively regulates push-pull reactions.
Due to this similarity, the $\cp$ kinetics may have robustness for the detection of the ligand concentration in a noisy environment.

More realistically, T cells, for example, have to detect their target ligands among a large number of non-target ligands.
Therefore, to understand the discrimination mechanism in the presence of a high concentration of non-target ligands is also important. 
In this work, we did not deal with the situation that a cell discriminates ligands in the mixture of several ligands.
Meanwhile, several theoretical researches have considered the situation of a mixture of ligands in the stochastic condition \cite{2016JSP...162.1130F, 2013PhRvL.110u8102L, 2015PhRvL.115c8102M}. 
The above push-pull kinetics regulated by a bifunctional molecule in \Ref\cite{2015PhRvL.115c8102M} is proposed as a ligand discrimination model in the ligand mixture condition, suggesting that the $\cp$ kinetics, which also has a bifunctional molecule, may discriminate ligands in the mixture condition.
However, these researches only consider the mixture condition where there are two types of ligands, target and non-target. 
In a real biological system, a cell detects its target ligands against the background of various types of non-target ligands.
In the realistic condition, the effectiveness of the mechanisms or kinetics proposed for ligand discrimination in which a cell discriminates target from non-target in the presence of a mixture of the two types of ligands is not obvious.
Further study is needed to understand the mechanisms of ligand discrimination in the presence of various types of ligands.

In this work, we employ a mechanism of nonlinear dynamics in a biological system and relate it to the ligand discrimination problem by using several kinetic motifs. 
Our findings are not restricted to the concrete kinetics we have shown in this work and have the potential to be employed for various other systems for ligand discrimination. 
For example, the mathematical analysis of our simple kinetics for ligand discrimination demonstrates the mathematically fundamental mechanisms of these kinetics. 
Beyond the kinetic motifs, the mathematical mechanisms provide a perspective on various theoretical models and experimental data in a real biological ligand discrimination system. 

In order to reach a further understanding of the ligand discrimination system, we have to deal with the problems of the stochastic effect and the existence of various types of non-target ligands. 
However, achieving a fundamental understanding of deterministic and simple systems will help to understand the underlying mechanisms of the stochastic and more complex ligand discrimination system. 